\documentclass[11pt,english,a4paper,floatfix]{article}
\usepackage{jcappub}
\usepackage{euclid}

\usepackage{soul}

\usepackage{scalerel,tikz}
\usetikzlibrary{svg.path}
\definecolor{orcidlogocol}{HTML}{A6CE39}
\tikzset{orcidlogo/.pic={
 \fill[orcidlogocol] svg{M256,128c0,70.7-57.3,128-128,128C57.3,256,0,198.7,0,128C0,57.3,57.3,0,128,0C198.7,0,256,57.3,256,128z};
 \fill[white] svg{M86.3,186.2H70.9V79.1h15.4v48.4V186.2z}
 svg{M108.9,79.1h41.6c39.6,0,57,28.3,57,53.6c0,27.5-21.5,53.6-56.8,53.6h-41.8V79.1z M124.3,172.4h24.5c34.9,0,42.9-26.5,42.9-39.7c0-21.5-13.7-39.7-43.7-39.7h-23.7V172.4z}
 svg{M88.7,56.8c0,5.5-4.5,10.1-10.1,10.1c-5.6,0-10.1-4.6-10.1-10.1c0-5.6,4.5-10.1,10.1-10.1C84.2,46.7,88.7,51.3,88.7,56.8z};
}}
\newcommand{\orcid}[1]{\,\href{https://orcid.org/#1}{\mbox{\scalerel*{
\begin{tikzpicture}[yscale=-1,transform shape]
\pic{orcidlogo};
\end{tikzpicture}
}{O}}}}

\pdfoutput=1
\usepackage{bm}
\usepackage{amsfonts}
\usepackage{subfigure}

\renewcommand{\vec}[1]{\boldsymbol{#1}}

\usepackage{lmodern}
\usepackage{slantsc}
\newcommand{\AREPO}{\texttt{AREPO}\xspace}
\newcommand{\CONCEPT}{\texttt{CO\textit{N}\hspace{-0.04em}CEPT}\xspace}
\newcommand{\CLASS}{\texttt{CLASS}\xspace}
\newcommand{\CAMB}{\texttt{CAMB}\xspace}
\newcommand{\GADGET}[1]{\texttt{GADGET-#1}}
\newcommand{\LGADGET}{\texttt{L-GADGET3}\xspace}
\newcommand{\openGADGET}{\texttt{openGADGET3}\xspace}
\newcommand{\PKDGRAV}{\texttt{PKDGRAV3}\xspace}
\newcommand{\RAMSES}{\texttt{RAMSES}\xspace}

\newcommand{\BACCOemulator}{\texttt{BACCOemulator}\xspace}
\newcommand{\SWIFT}{\texttt{SWIFT}\xspace}
\newcommand{\ANUBIS}{\texttt{ANUBIS}\xspace}
\newcommand{\COLA}{\texttt{COLA}\xspace}
\newcommand{\gevolution}{\texttt{gevolution}\xspace}
\newcommand{\PINOCCHIO}{\texttt{PINOCCHIO}\xspace}
\newcommand{\FastDF}{\texttt{FastDF}\xspace}
\newcommand{\EuclidEmulator}{\texttt{EuclidEmulator2}\xspace}
\newcommand{\CosmicEmu}{\texttt{Cosmic\,Emu}\xspace}
\newcommand{\ReACT}{\texttt{ReACT}\xspace}
\newcommand{\NMGADGET}{\texttt{NM-GADGET4}\xspace}
\newcommand{\Pylians}{\texttt{Pylians3}\xspace}
\newcommand{\halofit}{\textit{halofit}\xspace}
\newcommand{\HMcode}{\texttt{HMcode}\xspace}

\newcommand{\cb}{\text{cb}}
\newcommand{\Nyq}{\text{Nyq}}

\newcommand{\kv}{\vec{k}}
\newcommand{\vk}{\vec{k}}

\newcommand{\qv}{\vec{q}}

\begin{document}


\renewcommand{\thefootnote}{\fnsymbol{footnote}}

\title{\textit{Euclid}: Modelling massive neutrinos in cosmology -- a code comparison\footnote{This paper is published on behalf of the Euclid Consortium}}

\author[1]{J.~Adamek\orcid{0000-0002-0723-6740},}
\author[2,3]{R.~E.~Angulo\orcid{0000-0003-2953-3970},}
\author[4]{C.~Arnold,}
\author[5,6,7]{M.~Baldi\orcid{0000-0003-4145-1943},}
\author[8,9,10,11]{M.~Biagetti\orcid{0000-0002-5097-479X},}
\author[12,13,1]{B.~Bose\orcid{0000-0003-1965-8614},}
\author[14]{C.~Carbone,}
\author[9,10,8]{T.~Castro\orcid{0000-0002-6292-3228},}
\author[1,15]{J.~Dakin\orcid{0000-0002-2915-0315},}
\author[16,17]{K.~Dolag,}
% ...

\affiliation[1]{Institute for Computational Science, Universit\"at Z\"urich, Winterthurerstrasse 190, 8057 Z\"urich, Switzerland}
\affiliation[2]{Donostia International Physics Center (DIPC), Paseo Manuel de Lardizabal 4, 20018 Donostia-San Sebasti\'{a}n, Guipuzkoa, Spain}
\affiliation[3]{IKERBASQUE, Basque Foundation for Science, 48013 Bilbao, Spain}
\affiliation[4]{Department of Physics, Institute for Computational Cosmology, Durham University, South Road, Durham DH1 3LE, UK}
\affiliation[5]{Dipartimento di Fisica e Astronomia ``Augusto Righi'' -- Alma Mater Studiorum Universit\`{a} di Bologna, via Piero Gobetti 93/2, 40129 Bologna, Italy}
\affiliation[6]{INAF-Osservatorio di Astrofisica e Scienza dello Spazio di Bologna, Via Piero Gobetti 93/3, 40129 Bologna, Italy}
\affiliation[7]{INFN-Sezione di Bologna, Viale Berti Pichat 6/2, 40127 Bologna, Italy}
\affiliation[8]{IFPU, Institute for Fundamental Physics of the Universe, via Beirut 2, 34151 Trieste, Italy}
\affiliation[9]{INAF-Osservatorio Astronomico di Trieste, Via G.\ B.\ Tiepolo 11, 34143 Trieste, Italy}
\affiliation[10]{INFN, Sezione di Trieste, Via Valerio 2, 34127 Trieste, Italy}
\affiliation[11]{SISSA, International School for Advanced Studies, Via Bonomea 265, 34136 Trieste, Italy}
\affiliation[12]{Institute for Astronomy, University of Edinburgh, Royal Observatory, Blackford Hill, Edinburgh EH9 3HJ, UK}
\affiliation[13]{Universit\'e de Gen\`eve, D\'epartement de Physique Th\'eorique and Centre for Astroparticle Physics, 24 quai Ernest-Ansermet, 1211 Gen\`eve 4, Switzerland}
\affiliation[14]{INAF-IASF Milano, Via Alfonso Corti 12, 20133 Milano, Italy}
\affiliation[15]{Department of Physics and Astronomy, University of Aarhus, Ny Munkegade 120, 8000 Aarhus C, Denmark}
\affiliation[16]{Universit\"ats-Sternwarte M\"unchen, Fakult\"at f\"ur Physik, Ludwig-Maximilians-Universit\"at M\"unchen, Scheinerstrasse 1, 81679 M\"unchen, Germany}
\affiliation[17]{Max-Planck-Institut f\"ur Astrophysik, Karl-Schwarzschild Str.\ 1, 85741 Garching, Germany}
% ...

\emailAdd{julian.adamek@uzh.ch}

\collaborationImg{\includegraphics[width=6cm]{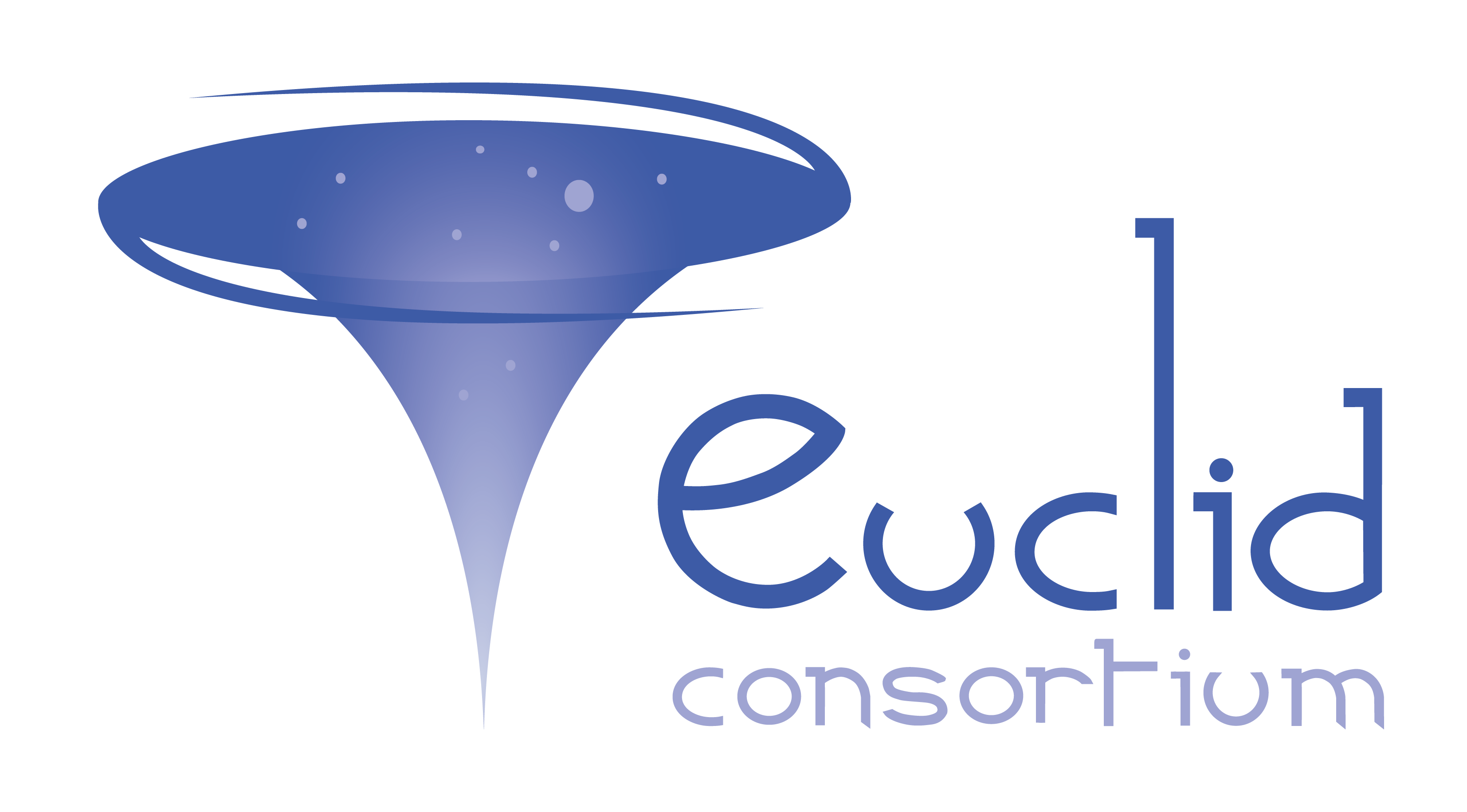}}

\abstract{
The measurement of the absolute neutrino mass scale from cosmological large-scale clustering data is one of the key science goals of the \textit{Euclid} mission. Such a measurement relies on precise modelling of the impact of neutrinos on structure formation, which can be studied with $N$-body simulations. Here we present the results from a major code comparison effort to establish the maturity and reliability of numerical methods for treating massive neutrinos. The comparison includes eleven full $N$-body implementations (not all of them independent), two $N$-body schemes with approximate time integration, and four additional codes that directly predict or emulate the matter power spectrum. Using a common set of initial data we quantify the relative agreement on the nonlinear power spectrum of cold dark matter and baryons and, for the $N$-body codes, also the relative agreement on the bispectrum, halo mass function, and halo bias. We find that the different numerical implementations produce fully consistent results. We can therefore be confident that we can model the impact of massive neutrinos at the sub-percent level in the most common summary statistics. We also provide a code validation pipeline for future reference.
}

\arxivnumber{2211.12457}

\toccontinuoustrue

\compress

\maketitle

\renewcommand{\thefootnote}{\arabic{footnote}}


\section{Introduction}
\label{sec:introduction}

The upcoming \textit{Euclid} mission \cite{EUCLID:2011zbd} will provide very detailed observations of the large-scale structure of our Universe, making it possible to probe physics related to dark energy and neutrinos at an unprecedented level of precision.
The analysis and interpretation of these data require a very accurate modelling of the process of structure formation. This is of particular relevance since a precise modelling of the mass-dependent effect neutrinos have on various summary statistics will allow a cosmological measurement of the absolute neutrino mass scale, which is one of the key science objectives of \textit{Euclid}.

Here we focus on the treatment of massive neutrinos in cosmological $N$-body simulations and investigate the convergence of a number of different codes over a variety of different scales and redshifts. Oscillation experiments have established a firm lower bound on the sum of neutrino masses of around $0.06\,\mathrm{eV}$. Using the well-known relation between neutrino mass and cosmological energy density, $\sum m_\nu \approx \Omega_\nu h^2 \times 94 \, \text{eV}$, this lower bound on the sum of neutrino masses corresponds to a lower bound of $\Omega_\nu h^2 > 6 \times 10^{-4}$, or approximately 0.5\% of the total matter density. As usual, the cosmological density of any component X can be given in terms of $\Omega_\text{X}$, which is its present-day energy density in units of the critical density, and physical density parameters are then denoted as $\omega_\text{X} = \Omega_\text{X} h^2$, where $h$ is the reduced Hubble parameter.

The sum of the neutrino masses is already constrained using information from the cosmic microwave background (CMB) combined with observations of large-scale structure like baryon acoustic oscillations \cite{DiValentino:2019dzu}, redshift-space distortions \cite{eBOSS:2020yzd}, and the Lyman-$\alpha$ forest \cite{Palanque-Delabrouille:2019iyz}. While currently providing only upper bounds, these constraints are expected to improve significantly with upcoming surveys like \textit{Euclid} which will be able to measure the neutrino mass fraction even if it is close to the lower bound. The matter power spectrum is still affected at the level of $4\,\%$ in this scenario, which is well within the sensitivity of the \textit{Euclid} main probes. 

For instance, the weak-lensing signal probed by \textit{Euclid} is sensitive to the matter power spectrum up to $k \approx 7\,h\,\mathrm{Mpc}^{-1}$ \cite{Taylor:2018nrc}. A linear model would be completely inadequate at such short scales and we therefore need robust nonlinear models. The forecast for galaxy clustering in \textit{Euclid} typically assumes that the matter power spectrum and the galaxy bias is well understood at least up to $k \approx 0.3\,h\,\mathrm{Mpc}^{-1}$ which also requires some nonlinear prescription \cite{Euclid:2019clj}. Many codes considered here are used in \textit{Euclid} preparation papers and reference simulations. For instance, \PKDGRAV ran \textit{Euclid}'s flagship simulations and the simulations used by Knabenhans et al.~\cite{Euclid:2018mlb,Euclid:2020rfv} to train the \EuclidEmulator; \openGADGET was used to calibrate the halo mass function for the \textit{Euclid} cluster abundance analysis by Castro et al.~\cite{Euclid:2022dbc}, and \PINOCCHIO was used to create the synthetic catalogues for the validation of the covariance matrix of cluster abundance and the clustering of clusters by Fumagalli et al.~\cite{Euclid:2021api,Euclid:2022txd}. Our objective is therefore to establish a reliable calibration baseline for the measurement of the neutrino mass scale within the cosmological analysis of \textit{Euclid} data. We expect that our results are also relevant in the context of other so-called ``stage IV surveys'' like the \textit{Vera C.\ Rubin Observatory} or the \textit{Nancy Grace Roman Space Telescope}.

Over the past decade, a variety of different methods have been employed to incorporate massive neutrinos in $N$-body simulations \cite{Angulo:2021kes}. As discussed in more detail later, they broadly fall into two categories which we may refer to as ``particle based'' and ``mesh based,'' respectively. They follow different philosophies of keeping track of the evolving neutrino phase-space distribution function. Particle-based methods sample the six-dimensional phase-space directly, see e.g.\ Refs.~\cite{Brandbyge:2008rv,Viel:2010bn,Agarwal:2010mt,Bird:2011rb,Villaescusa-Navarro:2013pva,Castorina:2015bma,Emberson:2016ecv,Adamek:2017uiq,Banerjee:2018bxy,Brandbyge:2018tvk}.
Mesh-based methods, on the other hand, work under the approximation that neutrino perturbations remain small and can be treated with perturbation theory
~\cite{Brandbyge:2008js,Ali-Haimoud:2012fzp,Liu:2017now,Chen:2020kxi}. In the simplest case that is sufficient for low neutrino masses one works with a linear realisation of the neutrino density field on a grid \cite{Brandbyge:2008js}. The linear theory for neutrinos may also be solved using the full nonlinear gravitational potential calculated in the simulation~\cite{Ali-Haimoud:2012fzp,Liu:2017now,Chen:2020kxi}. A different approach is to treat the massive neutrinos as a fluid and then solve the corresponding fluid equations, employing some approximation scheme to close the set of equations \cite{Dakin:2017idt}. One can also attempt to integrate the Vlasov--Poisson equations on a six-dimensional phase-space grid \cite{Yoshikawa:2020ehd}.

Generally speaking, the mesh-based schemes work best for relatively small neutrino masses where neutrino perturbations remain linear or quasi-linear at all times.
There are also hybrid schemes that use elements from both approaches. For instance, one may use a linear mesh-based representation at early times which is then converted to a particle representation at late time \cite{Brandbyge:2009ce,Bird:2018all}. The so-called ``$\delta f$ method'' introduced in Elbers et al.~\cite{Elbers:2020lbn} is another hybrid approach that uses a particle ensemble to estimate perturbations $\delta f$ to the smooth background phase-space distribution function $f$. Finally, a coordinate (gauge) transformation can be used to include linear neutrino perturbations without modifications to the $N$-body simulation code \cite{Partmann:2020qzb}. Cosmology-rescaling algorithms that are applied in post-processing have been shown to provide accurate results as well \cite{Zennaro:2019aoi}.
In this work, we aim to compare these different numerical approaches by employing them to run cosmological $N$-body simulations starting from the same initial conditions, and comparing the properties of the resulting matter and halo distributions using a controlled post-processing pipeline.
 
This paper is structured as follows. We begin with a brief review of neutrino physics and its impact on cosmology in Section \ref{sec:physics}. In Section \ref{sec:methods}, we describe the numerical methods that can be used to account for the cosmological effects of neutrinos. Our simulations are described in Section \ref{sec:simulations} and in Section \ref{sec:results} we present our numerical results. We conclude with a discussion in Section \ref{sec:discussion}.

\section{Neutrino physics}
\label{sec:physics}

From oscillation experiments it is firmly established that at least two of the standard-model neutrino mass states have non-zero mass, but the absolute mass scale is unknown and two mass orderings remain possible: normal and inverted. The current best-fit values for the mass-square differences measured in oscillation experiments are given by \cite{Esteban:2020cvm} 
\begin{eqnarray} 
 \Delta m_{21}^2 & = & 7.42^{+0.21}_{-0.20}\times 10^{-5} \, {\rm eV}^2, \\
  \Delta m_{31}^2 & = & 2.517^{+0.026}_{-0.028}\times 10^{-3} \, {\rm eV}^2 \quad (\textrm{NO}), \\
  \Delta m_{32}^2 & = & - 2.498^{+0.028}_{-0.028}\times 10^{-3} \, {\rm eV}^2 \quad (\textrm{IO}),   
\end{eqnarray}
where ``NO'' denotes the normal mass ordering and ``IO'' the inverted one. This leads to lower bounds on the sum of neutrino masses of $\sum m_\nu \gtrsim 0.06$ eV (NO) and $\sum m_\nu \gtrsim 0.1$ eV (IO), respectively. The best current experimental upper bound on the neutrino masses comes from the \textit{KATRIN} experiment, which measures an incoherent sum of mass states using beta decay of tritium \cite{KATRIN:2019yun,Aker:2021gma}. This bound approximately translates to $\sum m_\nu\lesssim 2.4$\,eV.

However, cosmology already provides much more stringent bounds, typically around 
$\sum m_\nu \lesssim 0.1-0.2$ eV. Assuming a minimal $\Lambda$CDM cosmology with massive neutrinos, a joint analysis of cosmological probes currently obtains $\sum m_\nu < 0.09$ eV at $95\%$ confidence, already putting pressure on the inverted mass ordering scenario \cite{DiValentino:2021hoh}.
These constraints are based on several physical effects affecting the CMB and large-scale structure in different ways, see Archidiacono et al.~\cite{Archidiacono:2016lnv} for a review. Some of the effects are simply related to the change in the expansion history, others to the explicit coupling of neutrinos to cosmological perturbations. More specifically,
massive neutrinos modify the shape of the matter power spectrum both in the linear and the nonlinear regimes.
First, as neutrinos behave like radiation in the early Universe, they move the radiation-matter equality to slightly later times, therefore shifting the peak of the power spectrum towards smaller wavenumbers.
Second, after the non-relativistic transition, they slow down the linear growth of perturbations at scales smaller than the free-streaming length, leading to a scale-dependent growth rate.
The small-scale suppression in the linear power spectra of cold dark matter (CDM) and baryons, $P_\mathrm{cb}$, or total matter (which includes massive neutrinos), $P_\mathrm{m}$, with respect to a model with massless neutrinos, can be quantified as \cite{Castorina:2015bma,Bond:1980ha,Hu:1997mj}
\begin{equation}
\frac{\Delta P_\mathrm{cb}}{P_\mathrm{cb}}\approx 6\, f_\nu\,,
\qquad
\frac{\Delta P_\mathrm{m}}{P_\mathrm{m}}\approx 8\, f_\nu\,,
\label{eq:pk_supp_cdm_baryons}
\end{equation}
respectively, where $f_\nu = \Omega_\nu / \Omega_\mathrm{m}$ is the neutrino mass fraction.
In the nonlinear regime, this suppression is even more prominent and exhibits a dip at $k\approx 1 \ h \ \mathrm{Mpc}^{-1}$ for low redshift, giving rise to the well-known ``spoon-like’’ feature \cite{Hannestad:2020rzl}.
In the context of the halo model, this feature appears at the transition region of the two-halo and the one-halo terms.
In particular, the dip is caused by the small-scale suppression of two-halo clustering that is induced by free-streaming, while the subsequent rise reflects the fact that the number of the most massive halos is rather independent of the neutrino masses.
All these effects can be accurately predicted by modelling the neutrino component in cosmological $N$-body simulations.
Assessing the relative accuracy and convergence of such modelling over a range of different numerical methods and simulation codes is the main goal of this paper.

\section{Numerical methods}
\label{sec:methods}

In this section, we give an overview of the various methods that have been developed for the treatment of massive neutrinos in $N$-body simulations and other numerical models. The most accurate results are expected when the local density of neutrinos in configuration space is accounted for within a simulation itself. This is technically challenging because of the large phase-space volume that is populated by neutrino particles. The methods to deal with this broadly fall into two categories that shall be discussed in turn: particle-based and mesh-based. Hybrid methods that use concepts from both categories have also been developed. Apart from full $N$-body simulations that may try to incorporate (as much as possible) the neutrino physics, there also exist approximate methods to generate realisations of large-scale structure. These can be augmented with recipes to account for the effect of massive neutrinos. Finally, if one is only interested in summary statistics like the power spectrum, emulators are a powerful tool that can be calibrated to include the sum of the neutrino masses as a free parameter. An overview of the various numerical codes employed in this work is given in Table \ref{tab:methods}.

\begin{table}
    \centering
    \caption{Overview of the numerical codes employed in this code comparison. $N$-body codes typically use a particle-mesh (PM) method coupled to some scheme to increase the force resolution in high-density regions. Methods featured here include Tree-PM, adaptive mesh refinement (AMR), fast multipole method (FMM), particle-particle/particle-mesh (P$^3$M), and moving mesh. No adaptive force computation is used in \gevolution, \COLA, and \PINOCCHIO, which all work with a uniform mesh.}
    \label{tab:methods}
    \bigskip
    \begin{tabular}{l||l|l|l}
       Code & type & neutrino method(s) & reference(s) \\
 \hline
       \GADGET{3} & $N$-body (Tree-PM) & particle & \cite{Springel:2005mi,Springel:2008cc} \\
       \LGADGET & $N$-body (Tree-PM) & mesh & \cite{Springel:2008cc,Angulo:2012ep,Angulo:2020vky} \\
       \openGADGET & $N$-body (Tree-PM) & particle &  \cite{Beck:2015qva,Marin-Gilabert:2022ggx} \\
       \GADGET{4} & $N$-body (Tree-PM) & particle & \cite{Springel:2020plp} \\
       \NMGADGET & $N$-body (Tree-PM) & Newtonian motion gauge & \cite{Partmann:2020qzb, Heuschling:2022rae} \\
       \AREPO & $N$-body (Tree-PM) & particle &  \cite{Springel:2009aa,Weinberger:2019tbd} \\
       \CONCEPT & $N$-body (P$^3$M) & mesh & \cite{Dakin:2021ivb,Dakin:2017idt} \\
       \PKDGRAV & $N$-body (Tree + FMM) & mesh & \cite{Potter:2016ttn} \\
       \SWIFT & $N$-body (PM + FMM) & particle / $\delta f$ & \cite{Schaller:2018,Elbers:2020lbn,Elbers:2022xid} \\
       \ANUBIS & $N$-body (PM + AMR) & particle & \cite{Teyssier:2001cp,Mauland:2023eax} \\
       \gevolution & $N$-body (uniform PM) & particle / mesh & \cite{Adamek:2017uiq,Adamek:2015eda,Adamek:2016zes} \\
       \COLA & $N$-body surrogate & mesh & \cite{Winther:2017jof,Wright:2017dkw} \\
       \PINOCCHIO & $N$-body surrogate & linear growth factor & \cite{Monaco:2013qta, Rizzo:2016mdr} \\
       \ReACT & $P(k)$ prediction & halo-model reaction & \cite{Cataneo:2019fjp,Bose:2021mkz} \\
       \BACCOemulator & $P(k)$ prediction & emulation & \cite{Angulo:2020vky} \\
       \EuclidEmulator & $P(k)$ prediction & emulation & \cite{Euclid:2020rfv} \\
       \CosmicEmu & $P(k)$ prediction & emulation & \cite{Lawrence:2017ost,Moran:2022iwe}
    \end{tabular}
\end{table}

\subsection{Particle-based methods}
\label{subsec:particlemethods}

Conceptually, the most straightforward method of accounting for cosmic neutrinos in a simulation is to represent them by a separate $N$-body ensemble. However, this method faces several challenges that need to be addressed carefully.

The first challenge is posed by the aforementioned phase-space volume that needs to be sampled. The phase-space distribution of neutrino particles typically has a very large velocity dispersion that is orders of magnitude larger than the bulk velocity. Therefore, representing the neutrinos with a small number of $N$-body particles that simply track the bulk velocities, which is essentially the method of choice for CDM or baryons, would completely miss the fact that most neutrinos are unbound and easily free stream out of gravity wells. Thus, the common practice is to sample the $N$-body particles from the true phase-space distribution, effectively performing a Monte-Carlo integration of the evolution equations. The main drawback of this method is that a poor sampling usually introduces significant shot noise, while high sampling rates quickly become very expensive as both the memory requirement and computations become completely dominated by the neutrino particle load. This is undesirable since neutrinos are just a tiny fraction of the matter after all.

Shot noise affects all moments of the distribution function, and in particular the density. This means that shot noise will also propagate into the gravitational field computed from the neutrino perturbations. While this could in principle severely degrade the accuracy of the gravitational evolution as a whole, the impact is actually mitigated by the fact that neutrinos only account for a very small fraction of the total matter, and the gravitational fields are therefore dominated by cold species. Nevertheless, some shot noise does propagate into the other matter species, particularly on large scales where the contribution of neutrino perturbations is largest. Various strategies have been developed to reduce the impact of shot noise, e.g. by filtering small-scale fluctuations. One effective strategy is to implement a statistical weighting of the neutrino particles, as is done in the $\delta f$ method \cite{Elbers:2020lbn}. This method works by decomposing the distribution function $f$ into an analytical background component $\bar{f}$ and a perturbation $\delta f$ computed from the particle ensemble. The weights are given at each time step by requiring phase-space density conservation. These weights are typically negligible, except for particles that are significantly perturbed, such as those captured by halos. Shot noise is thereby minimised as particles only contribute to the gravitational potential when needed. The $\delta f$ method has been implemented in a number of codes, but is only used by \SWIFT{} in this comparison (see Table~\ref{tab:methods}). Other strategies aimed at reducing shot noise while still using particles include alternate sampling of neutrino momenta \cite{Banerjee:2018bxy} and various hybrid methods \cite{Brandbyge:2009ce,Banerjee:2016zaa,Bird:2018all}.

The second challenge pertains to the kinematics of neutrino $N$-body particles. If one were to apply a similar time-stepping criterion as for cold matter species, the high velocities would typically result in extremely small integration time steps, making the simulations considerably more expensive. This is often solved by relaxing the time-stepping criterion and allowing the neutrino particles to travel a larger distance in each integration step than what would be considered ``safe'' for cold species. High-velocity neutrino particles then respond poorly to small-scale fluctuations in the gravitational forces. Given that the neutrino distribution at small scales is plagued by the shot-noise problem anyway, this additional problem is often considered to be of little concern.

Still regarding kinematics, further issues arise due to the relativistic nature of neutrinos. In the weak-field limit, the propagation of a collisionless massive particle is governed by the Hamiltonian equations of motion \cite{Ma:1993xs}
\begin{eqnarray}
\label{eq:relativistic_v}
 \vec{v} &=& \frac{\vec{p} c}{\sqrt{\vec{p}^2 + m^2 c^2 a^2}}\,,\\
\label{eq:relativistic_p}
 \vec{p}' &=& -\frac{2\vec{p}^2 + m^2 c^2 a^2}{c \sqrt{\vec{p}^2 + m^2 c^2 a^2}} \nabla \Psi\,,
\end{eqnarray}
where $\vec{v}$ is the peculiar velocity, $\vec{p}$ is the canonical momentum, $m$ is the particle's rest-mass, $a$ is the scale factor, and $\Psi$ is the gravitational potential, assuming as usual that gravitational slip can be neglected, i.e.\ that non-relativistic and ultra-relativistic particles essentially see the same potential. Here, a prime denotes the derivative with respect to conformal time and $c$ denotes the speed of light. The canonical momentum is conserved in the absence of a gravitational force. For non-relativistic particles one usually considers the limit $\vec{p}^2 \ll m^2 c^2 a^2$ in which the equations simplify to
\begin{eqnarray}
 \vec{v} &=& \frac{\vec{p}}{m a}\,,\\
 \vec{p}' &=& -m a \nabla\Psi\,.
\end{eqnarray}
This simpler set of equations has the advantage that it is easy to find integration methods that are symplectic, i.e.\ that preserve the phase-space volume exactly as demanded by Hamiltonian time evolution. Note also that, in the absence of a gravitational force, the peculiar velocity scales exactly as $\propto a^{-1}$ in this case.

Even though the evolution of high-momentum particles suffers from severe errors, including the breakdown of causality for $\vec{p}^2 > m^2 c^2 a^2$, some implementations might still use the simplified equations. The propagation of these errors into the clustering amplitude of matter is limited by the fact that high-momentum particles barely contribute to clustering in the first place. However, using the more accurate Eqs.~\eqref{eq:relativistic_v} and \eqref{eq:relativistic_p} is a more common choice. The issue of symplectic time integration in this case is discussed e.g.\ in Appendix A of Adamek et al.~\cite{Adamek:2017uiq} and Appendix D of Elbers et al.~\cite{Elbers:2020lbn}.

\subsection{Mesh-based methods}

The main alternative to particle-based methods is to represent the distribution function on a spatial mesh. Yoshikawa et al.~\cite{Yoshikawa:2020ehd} discretise the distribution function on a six-dimensional mesh in phase-space, 
however, a brute-force approach like this is rather expensive and cannot easily be applied to very large simulations where memory requirements are a particular concern. A possible way out is to take moments of the distribution function (density, bulk velocity, and so on) where the momentum coordinates are integrated out so that a discretisation in the three spatial dimensions is sufficient.

Experience from solving the hierarchy of moments (the so-called Boltzmann hierarchy) in linear perturbation theory shows that a considerable number of moments must be taken into account in order to reach good accuracy for the evolution of the lowest moments. This concerns in particular the density that also affects the clustering of other matter components through gravitational coupling. On the other hand, the numerical solutions are readily available in the linear regime where they can be expressed in terms of linear transfer functions. Throughout this paper, we follow the convention from standard linear cosmological perturbation theory where the transfer function $T_X$ of any perturbation variable $X$ in Fourier space is defined through the relation
\begin{equation}
    X(\kv,z) = T_X(k,z) \zeta(\kv)\,,
\end{equation}
where $\zeta(\kv)$ is the comoving curvature perturbation of the mode $\kv$ before it enters the horizon. Those transfer functions provide a deterministic factor by which any given initial random perturbation mode needs to be multiplied in order to obtain, for instance, the density perturbation at any given time. Using these transfer functions that can be calculated at the outset, a simulation code can therefore construct the linear density field of neutrinos at any point in time for any given realisation of the random initial conditions. This is precisely what basic mesh-based methods do: they use the density field of neutrinos extrapolated from linear perturbation theory, which is often a reasonable approximation because neutrinos do not cluster strongly. The method is free of shot noise and is by construction exact in the limit of linear perturbations. However, it obviously lacks any response to nonlinear gravitational potentials that develop in a simulation.

While the linear method produces results that are sufficiently accurate for many purposes, some more advanced approaches have been developed in attempts to address the shortcomings. Ali-Ha\"imoud \& Bird~\cite{Ali-Haimoud:2012fzp} solve for the transfer function of the neutrino perturbations using the nonlinear matter power spectrum of the simulation to construct an effective source term, assuming that the phase correlation between neutrino and matter perturbations remain largely intact even at nonlinear scales. Dakin et al.~\cite{Dakin:2017idt} employ the coupled evolution equations for the lowest moments in their nonlinear form. Then, to avoid having to calculate a large Boltzmann hierarchy for every wavevector represented in the simulation, the hierarchy is truncated by assuming that a ``scaling'' holds approximately for ratios of higher moments, where the scaling coefficients are taken from linear theory. This approximation is then used to close the system of equations using only a small number of nonlinear moments. By construction this method agrees with the simpler method in the limit of linear perturbations. While the resulting nonlinear neutrino density is somewhat more realistic, the distribution function still has some residual errors that cannot easily be reduced without including further moments in the nonlinear computation.

\subsection{Approximations and other methods}

For some purposes, such as the computation of covariance matrices for different cosmological probes, it is useful to have methods for making cosmological predictions that are faster, although less accurate, than the traditional $N$-body methods discussed so far. Here we present two such methods that can be used as surrogates for $N$-body simulations: the COmoving Lagrangian Acceleration (\COLA) approach and the PINpointing Orbit-Crossing Collapsed HIerarchical Objects (\PINOCCHIO) approach. In both cases, a speed-up is achieved by drastically simplifying the time integration in the particle evolution. Finally, we also present a method that avoids the need to include any neutrino physics in the actual $N$-body simulation altogether, apart from in the background solution. This method employs the so-called \textit{Newtonian motion gauge} and can be used with virtually any numerical scheme that solves the Newtonian gravity problem.

\subsubsection{COLA}
\label{sec:cola}

The \COLA approach by Tassev et al.~\cite{Tassev:2013pn} produces fast, approximate simulations of cosmological structure formation. Essentially, instead of solving for a full particle trajectory $\vec{x}(t)$, in this approach we solve for the deviations of the full trajectory about the trajectory predicted by second-order Lagrangian perturbation theory (2LPT) $\delta \vec{x}(t) = \vec{x}(t) - \vec{x}_{\rm 2LPT}(t)$. Since the evolution of the particles on large scales will be very close to that predicted by 2LPT, we can decrease the number of time steps of the simulation to trade accuracy at small scales for overall simulation speed while maintaining good accuracy at large scales. For a large number of time steps, the method effectively converges to a standard PM $N$-body method.

Adding massive neutrinos to the \COLA method was described by Wright et al.~\cite{Wright:2017dkw}, which also included an implementation in the {\tt MG-PICOLA} simulation code\footnote{\url{https://github.com/HAWinther/MG-PICOLA-PUBLIC}.} by Winther et al.~\cite{Winther:2017jof}. This implementation was carried over to the \COLA\ solver within {\tt FML}\footnote{\url{https://github.com/HAWinther/FML/tree/master/FML/COLASolver}.} which succeeded {\tt MG-PICOLA}. It is this implementation of the \COLA solver within {\tt FML} that we use in this paper. 
These implementations rely on the linear mesh-based method described above, i.e.\ we use the density field of neutrinos extrapolated from linear perturbation theory on a mesh for the PM part. For the 2LPT part of the \COLA code, we make a further approximation to the 2LPT equation and use the $\Lambda$CDM kernel to speed up the computation.

To demonstrate the key advantage of \COLA over traditional $N$-body codes, we use only $50$ time steps linearly distributed in scale factor for the \COLA simulations in this paper. However, the \COLA method does not work well for simulations starting from high initial redshifts when using a relatively small number of time steps; for a discussion on how to optimise initial redshift and number of time steps in \COLA simulations see Sections 4.1 and 4.3 of Izard et al.~\cite{Izard:2015dja}. Therefore, we use a slightly modified\footnote{The modifications are to the order in which (pseudo-)random numbers are drawn for the phases and amplitudes, such that we now do the same as the version of {\tt N-GenIC} used for the other methods in this paper, see Section~\ref{subsec:ICs}.} version of {\tt FML}'s built-in generator of initial conditions to generate initial particle data at $z=19$ instead of $z=127$ as is described in Section \ref{subsec:ICs} and used for the other methods in this paper. In addition, we use the \CAMB Boltzmann solver by Lewis et al.~\cite{Lewis:1999bs,Lewis:2002ah} to generate the density transfer function for massive neutrinos.  
Finally, we note that for all \COLA simulations in this paper we use a force grid that is a factor of three finer than the mean inter-particle distance; for a thorough investigation of the impact of varying this factor in \COLA simulations see Section 4.4 of Izard et al.~\cite{Izard:2015dja}.

\subsubsection{PINOCCHIO}

The \PINOCCHIO code\footnote{\url{https://github.com/pigimonaco/Pinocchio}.} \cite{Monaco:2001jg,Monaco:2013qta,Munari:2016aut} is an approximate method to generate halo catalogues in a very small fraction
(of the order of $1/1000$) of the time taken by an equivalent $N$-body simulation. Starting from a linear density field generated in Lagrangian space over a regular grid, its main goal is to construct halo catalogues by predicting which particles will end up in dark matter halos. To achieve this goal the algorithm first smoothes the linear density on a grid of smoothing radii, then uses ellipsoidal collapse to compute the collapse time of each particle. In the second step, it proceeds to group the collapsed particles into halos, using an algorithm that mimics their hierarchical clustering and distinguishes between halos and filaments.

As opposed to the other $N$-body methods employed in this work, \PINOCCHIO
does not integrate particle orbits but places halos at their final position 
using a single 2LPT or 3LPT displacement. Indeed, once the displacement fields are averaged over the multi-stream region that corresponds to a dark matter halo, LPT is very effective in predicting halo positions \cite{Munari:2017aau}. Another difference is that \PINOCCHIO does not start from a set of displaced particles but generates the linear density field internally. Having implemented the same sequence for populating modes in $k$-space as the {\tt N-GenIC} code that is used for the purpose of generating initial data in this work (see Section \ref{subsec:ICs}), it can reproduce the same large-scale structure if the same seed for random numbers is provided.

The extension of \PINOCCHIO to massive neutrinos is presented by Rizzo et al.~\cite{Rizzo:2016mdr} and is based on the result of Castorina et al.~\cite{Castorina:2013wga,Castorina:2015bma}
that the nonlinear clustering of massive neutrinos is negligible. 
We use \CAMB to compute linear power spectra in massive neutrino cosmologies, and compute the scale-dependent growth rate of matter by taking ratios of power spectra of CDM and baryons (i.e.\ without neutrinos) at different times. We also adapt the code to incorporate a scale-dependent growth rate. With respect to the original implementation of Rizzo et al.~\cite{Rizzo:2016mdr}, which was limited to 2LPT, we extend here the computation to third order: as shown by Munari et al.~\cite{Munari:2016aut} this results in a significant improvement at mildly nonlinear scales.

 Although the code has been conceived to predict the properties of dark matter halos, it can produce a full nonlinear density field as follows: particles that do not belong to halos are moved to their final position using 3LPT, halo particles are distributed around their halo center of mass following a Navarro--Frenk--White (NFW) profile \cite{Navarro:1996gj} with Maxwellian velocity distributions. This allows us to construct snapshots like an $N$-body simulation, representing density fields that are far more accurate than a straight LPT implementation. 
Because we have only one type of particle, to compute the power spectrum of CDM and baryons needed below (Section~\ref{sec:results}) we subtract the linear neutrino contribution from the total matter power spectrum obtained from the snapshot. To this end we also assume that the neutrino-matter cross-power spectrum, $P_{\nu,\mathrm{m}}(k)$, can be approximated by $P_{\nu,\mathrm{m}}(k) = \sqrt{P_\nu^\mathrm{L}(k) P_\mathrm{m}(k)}$, where the superscript ``L'' denotes a power spectrum from linear theory. This approximation  is strictly only true in the linear regime but we apply it at all scales. 

We do not expect \PINOCCHIO to be competitive with $N$-body codes in predicting the matter power spectrum: taking it as a sophisticated implementation of 3LPT, we expect it to lose power on scales smaller than $k=0.3\,h\,\rm{Mpc}^{-1}$ for the halo power spectrum and $k=0.2\,h\,\rm{Mpc}^{-1}$ for the matter power spectrum. It will not be competitive with \COLA as well which, being a PM code, can converge to the solution (on scales larger than the mesh) if a sufficient number of time steps is used. This better accuracy comes at a higher cost in computing time, by approximately a factor of eight in the configuration used in this paper (see also Blot et al.\ \cite{Blot:2018oxk} for a similar benchmark), as well as in memory since our \COLA runs use a grid three times finer than the mean particle separation. The \PINOCCHIO code is widely used especially to characterise the covariance of galaxy clustering measurements, thanks to its low computational cost and its ability to generate halo catalogues.

\subsubsection{Newtonian motion gauge}

The Newtonian motion gauge approach for massive neutrinos was developed by Partmann et al.~\cite{Partmann:2020qzb} and Heuschling et al.~\cite{Heuschling:2022rae}. It allows for a simulation of nonlinear CDM in an ordinary Newtonian $N$-body simulation while accounting for the impact of linear neutrinos via a modification of the dark matter initial conditions and by employing a dynamically evolving coordinate system. 
The method is agnostic towards the implementation of the $N$-body simulation and for this paper we choose to employ \GADGET{4} \cite{Springel:2020plp}. However, any method solving Newtonian nonlinear gravity is compatible, even methods other than $N$-body simulations. We would like to stress that our method is exact in the weak-field limit of general relativity (see Fidler et al.~\cite{Fidler:2018bkg}) and therefore captures the full effect of linear neutrino perturbations on the nonlinear matter clustering. 

The Newtonian motion approach allows for a very simple inclusion of massive neutrinos, requiring only three additional steps applied to a simulation without any neutrinos.
First, we start from a set of ``back-scaled'' initial conditions based on the present-day power spectrum of CDM and baryons in the Newtonian motion coordinates, excluding neutrino perturbations. In contrast to other neutrino methods presented in this paper, these initial conditions do not assume a scale-dependent growth, i.e.\ the rescaling of the power spectrum is done using the scale-independent growth factor $D_{+}$. The residual effect of decaying modes due to neutrinos is included in the construction of the present-day matter power spectrum in the Newtonian motion coordinate system. This also has the added benefit that ordinary generators of initial conditions can be used without modifications, provided that the correctly back-scaled Newtonian motion gauge power spectrum is used. We then evolve the initial data with the Newtonian solver, taking into account the impact of the massive neutrinos on the background evolution via the Hubble rate. Finally, we obtain the output in the Newtonian motion gauge. To make it comparable to the output of other methods, we need to transform the result to the gauge employed therein (usually the ``$N$-boisson'' gauge \cite{Fidler:2018bkg}). This step is realised by a displacement field acting on the particle positions that is implemented in a similar way to how the initial conditions are set. The transformation accounts for the residual impact of neutrinos and other relativistic effects on the evolution of the CDM and baryon particles. However, by construction, the transformation vanishes exactly at $z=0$ (or another chosen target redshift) while it is in general very small at late times, for small neutrino masses and on small scales. Therefore, it can often be neglected as it will only lead to small corrections in which case the output of the simulation can be used as-is. 

For this work, we include only the first two steps, while omitting the final particle displacement. This leads to a small mismatch in the results shown for $z=1$ at large scales for the cases of the highest neutrino masses. By leaving this correction out, we demonstrate that for most neutrino masses, box sizes, and redshifts the method is already sufficiently accurate in its simplest form. For more details on the transformation we refer the reader to the original work by Partmann et al.~\cite{Partmann:2020qzb}.

\subsection{Halo-model reaction}

The halo model reaction approach provides the nonlinear corrections caused by massive neutrinos to a $\Lambda$CDM power spectrum through a ratio of halo-model predictions. Following Cataneo et al.~\cite{Cataneo:2018cic}, the nonlinear power spectrum is then given by
\begin{equation}
    P^{\rm NL}(k,z) = \mathcal{R}(k,z)  P_{\rm pseudo}^{\rm NL} (k,z) \, , \label{eq:nlps}
\end{equation}
where $\mathcal{R}(k,z)$ is the halo-model reaction and $P_{\rm pseudo}^{\rm NL}(k,z)$ is the nonlinear pseudo power spectrum. 

The pseudo spectrum is a nonlinear $\Lambda$CDM power spectrum but with the initial conditions tuned such that its linear clustering exactly matches the linear clustering in the non-standard cosmology at the target redshift. For example, if the non-standard physics introduces a simple rescaling of the linear clustering amplitude, one could just rescale the amplitude of any $\Lambda$CDM power spectrum to produce the pseudo spectrum. In the case of scale-dependent modifications, this becomes a bit trickier in practice. We approximate this quantity as in previous works by Cataneo et al.~\cite{Cataneo:2018cic,Cataneo:2019fjp} and pass the modified linear spectrum as produced by \CAMB to \HMcode developed by Mead et al.~\cite{Mead:2020vgs}, with $\Lambda$CDM presets, i.e.\ no baryonic feedback nor massive neutrinos. The benefit of using the pseudo rather than $\Lambda$CDM cosmology is that it guarantees the mass functions in target and pseudo cosmologies are similar as they have the same linear clustering. This produces a smoother transition between the two-halo and one-halo terms. This transition was one of the previous issues in calculating this nonlinear response using the halo model \cite{Cooray:2002dia,Cacciato:2008hm,Giocoli:2010dm}. 

Following Cataneo et al.~\cite{Cataneo:2019fjp}, the halo-model reaction for massive neutrinos is given by
\begin{equation}
    \mathcal{R}(k)=\frac{\left(1-f_{\nu}\right)^{2} P^{\mathrm{HM}}_\cb(k)+2 f_{\nu}\left(1-f_{\nu}\right) P^{\mathrm{HM}}_{\nu,\cb}(k)+f_{\nu}^{2} P^{\mathrm{L}}_{\nu}(k)}{P^{\mathrm{L}}_{\mathrm{m}}(k)+P^{\mathrm{1h}}_{\mathrm{pseudo}}(k)} \, ,
    \label{eq:reaction}
\end{equation}
with ``cb'' denoting the CDM and baryon component and ``$\nu$'' denoting massive neutrinos.  We include the effects of massive neutrinos at the linear level in the numerator via the weighted sum of the nonlinear halo-model (cb) spectrum and the massive neutrino linear spectrum \cite{Agarwal:2010mt}. The components of the reaction are
\begin{equation}
    P^{\mathrm{HM}}_{\nu,\cb}(k) \approx \sqrt{P^{\mathrm{HM}}_{\cb}(k) P^{\mathrm{L}}_{\nu}(k)} \, , 
\end{equation}
\begin{equation}
    P^{\mathrm{HM}}_{\cb}(k) = P^{\mathrm{L}}_{\cb}(k)+P^{\mathrm{1h}}_{\cb}(k) \, . \label{eq:1hcb} 
\end{equation}
Explicitly, the one-halo terms are given as integrals over the Fourier space halo density profile $u(k,M)$ and the halo mass function $n(M)$,
\begin{equation}
    P^{1 \mathrm{h}}_{\cb}(k)=\int \mathrm{d} \ln M \, n_{\mathrm{cb}}(M)\left(\frac{M}{\bar{\rho}_{\mathrm{cb}}}\right)^{2}|u_{\mathrm{cb}}(k, M)|^{2} \, , 
\end{equation}
\begin{equation}
    P^{1 \mathrm{h}}_{\mathrm{pseudo}}(k)=\int \mathrm{d} \ln M \, n_{\mathrm{pseudo}}(M)\left(\frac{M}{\bar{\rho}_{\mathrm{m}}}\right)^{2}|u_{\mathrm{pseudo}}(k, M)|^{2} \, , 
\end{equation}
where $\bar{\rho}$ is the background density for the specified matter species. The halo mass functions are given as 
\begin{equation}
    n_{\cb}(M) = \frac{\bar{\rho}_{\cb}}{M}[\nu' f(\nu')] \frac{\mathrm{d} \ln \nu'}{\mathrm{d} \ln M} \, ,
\end{equation}
\begin{equation}
    n_{\mathrm{pseudo}}(M) = \frac{\bar{\rho}_{\mathrm{m}}}{M}[\nu'' f(\nu'')] \frac{\mathrm{d} \ln \nu''}{\mathrm{d} \ln M} \, .
\end{equation}
The peak heights are  defined as $\nu' = \delta_{\mathrm{sc, cb}}(M)/\sigma_{\mathrm{cb}}[R_{\mathrm{cb}}(M)]$ and $\nu'' = \delta_{\mathrm{sc,m}}/\sigma_{\mathrm{m}}[R_{\mathrm{m}}(M)]$, where the subscript ``sc'' indicates this quantity is calculated by solving the standard $\Lambda$CDM spherical-collapse equations but with the indicated matter density. The mass fluctuation variances are given by 
\begin{equation}
    [\sigma_{\mathrm{cb}}(R)]^2 = \int \frac{\mathrm{d}^{3} k}{(2 \pi)^{3}}|\tilde{W}(k R)|^{2} P^{\mathrm{L}}_{\cb}(k),
\end{equation}
\begin{equation}
    [\sigma_{\mathrm{m}}(R)]^2 = \int \frac{\mathrm{d}^{3} k}{(2 \pi)^{3}}|\tilde{W}(k R)|^{2} P^{\mathrm{L}}_{\rm m}(k).
\end{equation}
In all predictions from halo-model reaction, we employ a Sheth--Tormen halo mass function \cite{Sheth:1999mn,Sheth:2001dp}, a power-law concentration--mass relation (see for example the work by Bullock et al.~\cite{Bullock:1999he}), and an NFW halo density profile \cite{Navarro:1996gj}. The predictions are computed numerically using the public code \ReACT\footnote{\url{https://github.com/nebblu/ReACT}} by Bose et al.~\cite{Bose:2020wch,Bose:2021mkz}.

\subsection{Power-spectrum emulation}

Fast predictors of the matter power spectrum are an essential ingredient for many inference pipelines in cosmology. Since numerical simulations are too costly to be directly applied in this context, different approaches based on approximate methods or elaborate fitting techniques have been used in the past. Apart from the halo-model reaction discussed in the previous section, well-known examples are the \halofit{} predictor developed by Smith et al.~\cite{Smith:2002dz} and later improved by Takahashi et al.~\cite{Takahashi:2012em}, and the \HMcode{} predictor by Mead et al.~\cite{Mead:2020vgs}, the latter being based on the halo model. In terms of the power-suppression signal of neutrinos, fitting routines by Bird et al.~\cite{Bird:2011rb} have been used in the past.

More recently, the emulation technique has become a popular alternative to obtain fast predictions of the matter power spectrum within the cosmological parameter space. Broadly speaking, emulators are interpolation routines based on a suite of numerical simulations that sample the cosmological parameter space and act as a training set. There are different surrogate techniques currently used for cosmological emulators, such as Gaussian process regression \cite{Heitmann:2009cu,Lawrence:2017ost,Nishimichi:2018etk,Giri:2021qin}, polynomial chaos expansion \cite{Euclid:2018mlb,Euclid:2020rfv}, or neural network approaches \cite{Arico:2020lhq}.

In addition to the predictions from \halofit{} and \HMcode{} mentioned above, we focus in this paper on the \CosmicEmu \cite{Lawrence:2017ost}, the \EuclidEmulator \cite{Euclid:2020rfv}, and the \BACCOemulator \cite{Angulo:2020vky}. These emulators provide predictions of the matter power spectrum and include a free parameter for the sum of the neutrino masses. We will now summarise the particularities of these three emulators, specifically focusing on the neutrino implementation.
\begin{itemize}
    \item The \CosmicEmu-2022 is built upon the \textit{Mira-Titan} simulations \cite{Heitmann:2015xma,Moran:2022iwe}, a suite of 111 simulations run with the {\tt HACC} code \cite{Habib:2014uxa}. The simulations are distributed over an eight-dimensional cosmological parameter space comprising ($\omega_\mathrm{m}$, $\omega_\mathrm{b}$, $\omega_{\nu}$, $\sigma_8$, $h$, $n_\mathrm{s}$, $w_0$, $w_a$), where $\sigma_8$ is the present-day amplitude of linear matter density fluctuations at the scale of $8\,\si{\hMpc}$, $n_s$ is the scalar spectral index, and $w_0$, $w_a$ parameterise the effective equation of state of dark energy in terms of the first two coefficients of the Taylor series expansion around $a=1$. The emulator achieves an absolute precision of about four percent for modes of $k<5\,h\,\mathrm{Mpc}^{-1}$ within the redshift range $z\in [0, 2]$. Neutrinos are not incorporated in the simulations and are effectively treated as a smooth background component. The power spectra are then corrected on large scales for enhanced growth beyond the neutrino free-streaming scale using the scale-dependent linear growth factor.
    
    \item The \EuclidEmulator is trained on 200 paired and fixed simulations that were run with \PKDGRAV. It emulates the nonlinear boost factor that is then multiplied by the results of a linear Boltzmann solver. The emulator covers eight cosmological parameters ($\Omega_\mathrm{m}$, $\Omega_\mathrm{b}$, $\sum m_{\nu}$, $A_\mathrm{s}$, $h$, $n_\mathrm{s}$, $w_0$, $w_a$), where $A_s$ is the amplitude of primordial perturbations at the scale $k_\text{p} = 0.05\,\mathrm{Mpc}^{-1}$, and includes redshifts of $z\in [0, 3]$ and modes up to $k=10\,h\,\mathrm{Mpc}^{-1}$. It claims an error of below one percent which is better than the other emulators discussed here. Note, however, that the \EuclidEmulator covers a somewhat narrower parameter space motivated by the results of the \textit{Planck} mission \cite{Planck:2015fie}. Within the training set the neutrinos are modelled using the mesh-based method implemented in \PKDGRAV. 
    
    \item The \BACCOemulator is trained on a very large suite of simulations based on the cos\-mology-rescaling technique \cite{Angulo:2009rc}. More specifically, four high-resolution simulations with judiciously chosen cosmologies are rescaled to more than 800 cosmologies at different redshifts. Whenever the target cosmologies included massive neutrinos, their effect is added following the extension of the cosmology-rescaling technique presented Zennaro et al.~\cite{Zennaro:2019aoi}. This emulator varies eight cosmological parameters ($\Omega_\cb$, $\Omega_\mathrm{b}$, $\sum m_{\nu}$, $\sigma_8$, $h$, $n_\mathrm{s}$, $w_0$, $w_a$) and covers a redshift range of $z\in [0, 1.5]$ for modes up to $k=5\,h\,\mathrm{Mpc}^{-1}$. The claimed precision is better than three percent.
\end{itemize}
We refer to the original references for more information about the emulators.

\section{Simulations}
\label{sec:simulations}

To compare different numerical methods, we carry out a large suite of $N$-body simulations where we employ different codes to run the same set of ten simulations summarised in Table~\ref{tab:simulations}. These ten simulations cover different choices of total neutrino mass $\sum m_\nu$ (including the massless case), different box sizes $L_\mathrm{box}$, and different mass resolutions to check for numerical convergence with respect to finite-volume and discretisation effects. $N_\mathrm{part}$ denotes the number of particles used for CDM and baryons, as well as the number of particles for neutrinos if a particle-based method is employed. For simplicity, we assume degenerate neutrino mass eigenstates because cosmology is mainly sensitive to the total neutrino mass scale \cite{Archidiacono:2020dvx}. We keep the total matter density (at redshift zero) fixed at $\Omega_\mathrm{m} = 0.319$ by adjusting the CDM density parameter together with the neutrino mass. The baryon density is fixed at $\Omega_\mathrm{b} = 0.049$, and the remaining cosmological parameters are $A_\mathrm{s} = 2.215 \times 10^{-9}$ at the pivot scale $k_\mathrm{p} = 0.05\,\mathrm{Mpc}^{-1}$, $n_\mathrm{s} = 0.9619$, and $h = 0.67$, which are based on the \textit{Euclid Flagship 2} simulation. Dark energy is modelled as a cosmological constant that provides a spatially flat background solution, and the CMB temperature is set to $2.7255\,\rm{K}$ and the effect of radiation is taken into account in the simulations at the linear level, either by using carefully tailored initial conditions as detailed below or by including the radiation component on the mesh if a mesh-based method is used.

Using identical initial data (see Section~\ref{subsec:ICs} below for details) in each case, the ten simulations are run with each of the thirteen $N$-body methods listed in Table~\ref{tab:methods} to produce particle snapshots at redshifts $z = 1$ and $z = 0$. For \AREPO, due to resource constraints, only the four simulations with $N_\mathrm{part} = 512^3$ are run, precluding the possibility of conducting numerical convergence tests in this case. Therefore, a total of $248$ individual snapshots are analysed in this code comparison. In addition, nonlinear power spectra are predicted for each distinct choice of neutrino masses using the remaining methods in Table~\ref{tab:methods}, as well as using the \HMcode{} and \halofit fitting methods.

\begin{table}
    \centering
    \caption{Overview of the basic parameters used in our simulation suite. The cases with $\sum m_\nu = 0\,\mathrm{eV}$ and $\sum m_\nu = 0.15\,\mathrm{eV}$ are the two main baselines for our comparison, but we include some cases with larger masses, up to $\sum m_\nu = 0.6\,\mathrm{eV}$, to probe more ``extreme'' regions of parameter space.}
    \label{tab:simulations}
    \bigskip
    \begin{tabular}{l||c|c|c|c}
       Simulation  & $L_\mathrm{box}$ & $N_\mathrm{part}$ & mass resolution & $\sum m_\nu$ \\
 \hline
       \texttt{0.0eV}  & 512 \si{\hMpc} & 512${}^3$ & 8.85$\times$10${}^{10}~h^{-1} \si{\solarmass}$ & 0.0 eV \\
       \texttt{0.15eV}  & 512 \si{\hMpc} & 512${}^3$ & 8.75$\times$10${}^{10}~h^{-1} \si{\solarmass}$ & 0.15 eV \\
       \texttt{0.3eV}  & 512 \si{\hMpc} & 512${}^3$ & 8.65$\times$10${}^{10}~h^{-1} \si{\solarmass}$ & 0.3 eV \\
       \texttt{0.6eV}  & 512 \si{\hMpc} & 512${}^3$ & 8.45$\times$10${}^{10}~h^{-1} \si{\solarmass}$ & 0.6 eV \\
       \texttt{0.0eV\_HR}  & 512 \si{\hMpc} & 1024${}^3$ & 1.11$\times$10${}^{10}~h^{-1} \si{\solarmass}$ & 0.0 eV \\
       \texttt{0.15eV\_HR}  & 512 \si{\hMpc} & 1024${}^3$ & 1.09$\times$10${}^{10}~h^{-1} \si{\solarmass}$ & 0.15 eV \\
       \texttt{0.0eV\_1024Mpc}  & 1024 \si{\hMpc} & 1024${}^3$ & 8.85$\times$10${}^{10}~h^{-1} \si{\solarmass}$ & 0.0 eV \\
       \texttt{0.15eV\_1024Mpc}  & 1024 \si{\hMpc} & 1024${}^3$ & 8.75$\times$10${}^{10}~h^{-1} \si{\solarmass}$ & 0.15 eV \\
       \texttt{0.3eV\_1024Mpc}  & 1024 \si{\hMpc} & 1024${}^3$ & 8.65$\times$10${}^{10}~h^{-1} \si{\solarmass}$ & 0.3 eV \\
       \texttt{0.6eV\_1024Mpc}  & 1024 \si{\hMpc} & 1024${}^3$ & 8.45$\times$10${}^{10}~h^{-1} \si{\solarmass}$ & 0.6 eV
    \end{tabular}
\end{table}

\subsection{Initial conditions}
\label{subsec:ICs}

The initial conditions of all simulations are generated at redshift $z=127$.\footnote{\COLA exceptionally uses $z=19$ as initial redshift (see Section~\ref{sec:cola}).} The linear matter power spectra and transfer functions are obtained by running either \CAMB or the \CLASS Boltzmann code by Blas et al.~\cite{Blas:2011rf}. These files are then used by the \texttt{REPS}\footnote{\url{https://github.com/matteozennaro/reps}} code to compute the \textit{rescaled} power spectra and transfer functions at $z=127$ by solving the multi-fluid linear equations as outlined by Zennaro et al.~\cite{Zennaro:2016nqo}. This procedure, known as \textit{rescaling}, guarantees that the power spectrum of the output of the simulation on linear scales at low redshift will match the correct linear power spectra. 
A realisation of initial data is then generated by drawing random phases for all perturbation modes and fixing their amplitudes according to the initial transfer functions. This approach of ``fixing'' the amplitudes effectively removes cosmic variance at linear scales and has been shown to generally produce less noisy summary statistics \cite{Angulo:2016hjd}. It introduces a specific type of non-Gaussianity that is not expected to affect any of our results.
Given the density field of CDM and baryons, the initial positions and velocities of the $N$-body particles are computed from the Zeldovich approximation \cite{Zeldovich:1969sb}. We employ a modified version of the \texttt{N-GenIC} code\footnote{\url{https://github.com/franciscovillaescusa/N-GenIC_growth}} that accounts for the scale-dependence present in both the growth rate and growth factor in cosmologies with massive neutrinos.

For neutrinos, the different implementations make use of distinct methods.
For particle-based implementations, the positions and velocities of the massive neutrino particles are generated in a similar fashion as CDM. This means we effectively make use of the Zeldovich approximation to set the first two moments of the phase-space distribution function which correspond to the density perturbation and bulk velocity, respectively. The initial velocities of neutrino particles are then offset by a random thermal component drawn from their Fermi--Dirac distribution \cite{Davis:1992ui,Klypin:1992sf}. We assume the standard Big Bang scenario where this distribution is set in equilibrium before the weak interaction freezes out --- when the Universe was about 1 second old --- and after freeze-out simply redshifts as the Universe expands. Note that this means that the typical thermal velocities are much larger than those of a distribution that is in thermal equilibrium at low redshift. For \SWIFT, neutrino particles are instead set up with the \FastDF{} code,\footnote{\url{https://github.com/wullm/fastdf}} using geodesic integration from high redshift \cite{Ma:1993xs,Adamek:2017uiq}, to reproduce the full distribution function and to prevent the initial perturbations from being erased by thermal motions \cite{Elbers:2022xid}.
For mesh-based implementations, on the other hand, the density field of each neutrino species is directly computed using the phases from the random field realisation of the linear initial conditions.

\subsection{Post-processing pipeline}

In order to quantify differences in our numerical schemes as precisely as possible, we analyse the snapshots of all our $N$-body simulations in a common pipeline. We compute the power spectra and bispectra of the CDM and baryon component, and produce halo catalogues from which we measure the halo mass functions and halo bias. In cases where the simulations provide a neutrino distribution, we also compute the cross-power spectra of neutrinos with the CDM and baryon component, as well as the neutrino auto-power spectra.

\subsubsection{Power spectra}

The power spectra of the different snapshots have been estimated using \Pylians.\footnote{\url{https://pylians3.readthedocs.io}} The routine first deposits particle masses into a regular 3D grid with $N^3$ voxels using the cloud-in-cell mass-assignment scheme. In this work, we always use a mesh with $N^3 = N_\mathrm{part}$ such that the Nyquist scales match between particles and mesh. Although using larger grids may improve measurements on smaller scales, we recommend caution due to potential systematics and advise against relying on results near or beyond the Nyquist scale set by the mean particle separation. The constructed field is then Fourier transformed and the effects of the mass-assignment scheme are corrected. Next, for each mode the square of its amplitude is computed, $|\delta(\vec{k})|^2$. The modes are then binned in intervals of width equal to the fundamental frequency $k_\mathrm{F} = 2 \pi L_\mathrm{box}^{-1}$ and the power spectrum is finally estimated as
\begin{equation}
    P(k_i)=\frac{1}{N_i}\sum_{\vec{k}\in B_i} |\delta(\vec{k})|^2\,,
\end{equation}
where $N_i$ is the number of independent modes in the considered bin $B_i = \{\vec{k} \,|\, i\, k_\mathrm{F} \leq |\vec{k}| < (i+1) k_\mathrm{F}\}$
and $k_i$ is computed as 
\begin{equation}
    k_i = \frac{1}{N_i}\sum_{\vec{k}\in B_i} |\vec{k}|\,.
\end{equation}
To compute the cross-power spectrum of two fields instead, the estimator is generalised in the most straightforward way,
\begin{equation}
    P_{\mathrm{X},\mathrm{Y}}(k_i)=\frac{1}{2 N_i}\sum_{\vec{k}\in B_i} \Bigl[\delta_\mathrm{X}(\vec{k})\delta_\mathrm{Y}^\ast(\vec{k}) + \delta_\mathrm{X}^\ast(\vec{k})\delta_\mathrm{Y}(\vec{k})\Bigr]\,,
\end{equation}
where the subscripts ``X'' and ``Y'' denote the two different fields.
When presenting the results, we combine the measurements into larger bins logarithmically spaced in $k$. This reduces the noise at large $k$ and makes our plots more readable.

Some codes do not produce snapshots at exactly the desired redshift, but at redshifts that deviate by less than $\pm\,0.01$ from the target redshift. These differences can be visible when comparing power spectra on a sub-percent level. For those cases, we rescale the power spectra by the square of the ratio of the linear growth factors at the respective redshift values. 
Such a rescaling is applied to \ANUBIS{} for all snapshots and to \LGADGET and \PKDGRAV at $z=1$.

\subsubsection{Bispectra}

We measure the bispectrum of CDM and baryons (ccc) using the estimator
\begin{equation}
\label{eq:estB}
\hat{B}_\mathrm{ccc}(k_l,k_m,k_n)  \equiv  \frac{k_\mathrm{F}^3}{N_\mathrm{tr}(k_l,k_m,k_n)}\sum_{\qv_1 \in B_l}\sum_{\qv_2 \in B_m}\sum_{\qv_3 \in B_n}\,\delta_\mathrm{K}(\qv_1 +\qv_2+\qv_3)\, \,\delta(\qv_1)\,\delta(\qv_2)\,\delta(\qv_3),
\end{equation}
where $N_\mathrm{tr}$ is the number of ``fundamental triangles'',
\begin{equation}\label{eq:Nt}
N_\mathrm{tr}(k_l,k_m,k_n)  \equiv  \sum_{\qv_1 \in B_l}\sum_{\qv_2 \in B_m}\sum_{\qv_3 \in B_n}\,\delta_\mathrm{K}(\qv_1 +\qv_2+\qv_3)\,,
\end{equation}
formed by the vectors $\qv_i$ satisfying the triangle condition $\qv_1 +\qv_2+\qv_3=0$ that are included within the ``triangle bin'' defined by the triplet of centers $(k_l,k_m,k_n)$ and corresponding bins $B_l, B_m, B_n$. 

We use a Python code implementing the fourth-order density interpolation and the interlacing scheme described
by Sefusatti et al.~\cite{Sefusatti:2015aex}. 
In order to compare the large-volume simulations ($L_\mathrm{box} = 1024\,h^{-1}\,\mathrm{Mpc}$) more easily with the small-volume ones, we use the same $k$-space binning in both cases, fixing the bin width to $k_\mathrm{F}$ of the small box.
Just like for the power spectra, to account for inaccuracies in the redshift of some snapshots, we rescale some of the resulting bispectra by the cube of the ratio of the linear growth factors at the respective redshift values. 
Such a rescaling is applied to \ANUBIS{} for all snapshots and to \LGADGET and \PKDGRAV at $z=1$.

\subsubsection{Halo catalogues}

For the considered snapshots of the various simulations, we identify halos with the code
\texttt{Denhf} \cite{Tormen:1998fp,Tormen:2003kb,Giocoli:2007uv,Despali:2015yla} which uses a ``spherical overdensity'' criterion. The algorithm does not rely on any pre-identification method. Only CDM and baryon particles are considered in the characterisation of halos; neutrino particles (if present at all in the simulation) are considered as
a background component \cite{Villaescusa-Navarro:2013pva,Castorina:2013wga}. 

\texttt{Denhf} estimates the local 
density at the position of each $N$-body particle by calculating the distance to its 10$^{\rm th}$-nearest neighbour $d_{10}$, and assigning to each particle a density that is proportional 
to $d_{10}^{-3}$.
Centered on the particle with the highest density value,
the algorithm grows a sphere and stops
when the mean density within the sphere falls below a desired overdensity threshold, set to $200$ times the background density of CDM and baryons for the purpose of this work.
All particles assigned to this spherical overdensity halo are then removed from the global list of particles, and the algorithm proceeds recursively
until none of the remaining particles has a local density large enough to be the center of a 
10-particle halo. Particles not assigned to halos will be part of the field.

Only in the case of \PINOCCHIO we use the halo catalogues as produced by the code itself instead of \texttt{Denhf}. Because \PINOCCHIO is calibrated on the friends-of-friends halo mass function, we translate its masses to spherical overdensity ones by applying the rescaling of halo masses that translates the halo mass function of Watson et al.\ \cite{Watson:2012mt}, that has been used to calibrate the code, to the one of Tinker et al.\ \cite{Tinker:2010my}. Such a rescaling has been used, e.g., by Fumagalli et al.\ \cite{Euclid:2021api} to force the halo mass function averaged over 1000 realisations to follow a target one. We compute the rescaling only once, in the case of massless neutrinos, and use it for all neutrino masses.

For the codes that do not produce snapshots at the exact values of the desired redshifts, we apply no further corrections here. The error in the redshift is less than $\pm\,0.01$ while our halo properties typically display disagreements larger than 1\% between different codes. We therefore assume that the error due to mismatching redshift values is subdominant.

\section{Results}
\label{sec:results}

\subsection{Power spectra}

A key prediction from neutrino simulations is a suppression of the matter power spectrum that exceeds the maximum linear theory prediction of $\Delta P_\text{m}/P_\text{m}\approx 8f_\nu$. The matter power spectrum can be decomposed as follows
\begin{equation} \label{eq:Pm_contributions}
P_\text{m}(k) = (1-f_\nu)^2P_\cb(k) + 2f_\nu(1-f_\nu)P_{\nu,\cb}(k) + f_\nu^2P_\nu(k)\,,
\end{equation}
where $P_\cb$ is the power spectrum of CDM and baryons, $P_{\nu,\cb}$ the cross-power spectrum of neutrinos with CDM and baryons, and $P_\nu$ the neutrino power spectrum. Various methods treat these components differently or make predictions for only some of them, so we discuss each component in turn. Finally, we will also compare the total $P_\text{m}(k)$ with various power spectrum emulators, including \EuclidEmulator which only predicts this quantity.

\subsubsection{CDM and baryons}

\begin{figure}
  \includegraphics[width=\textwidth]{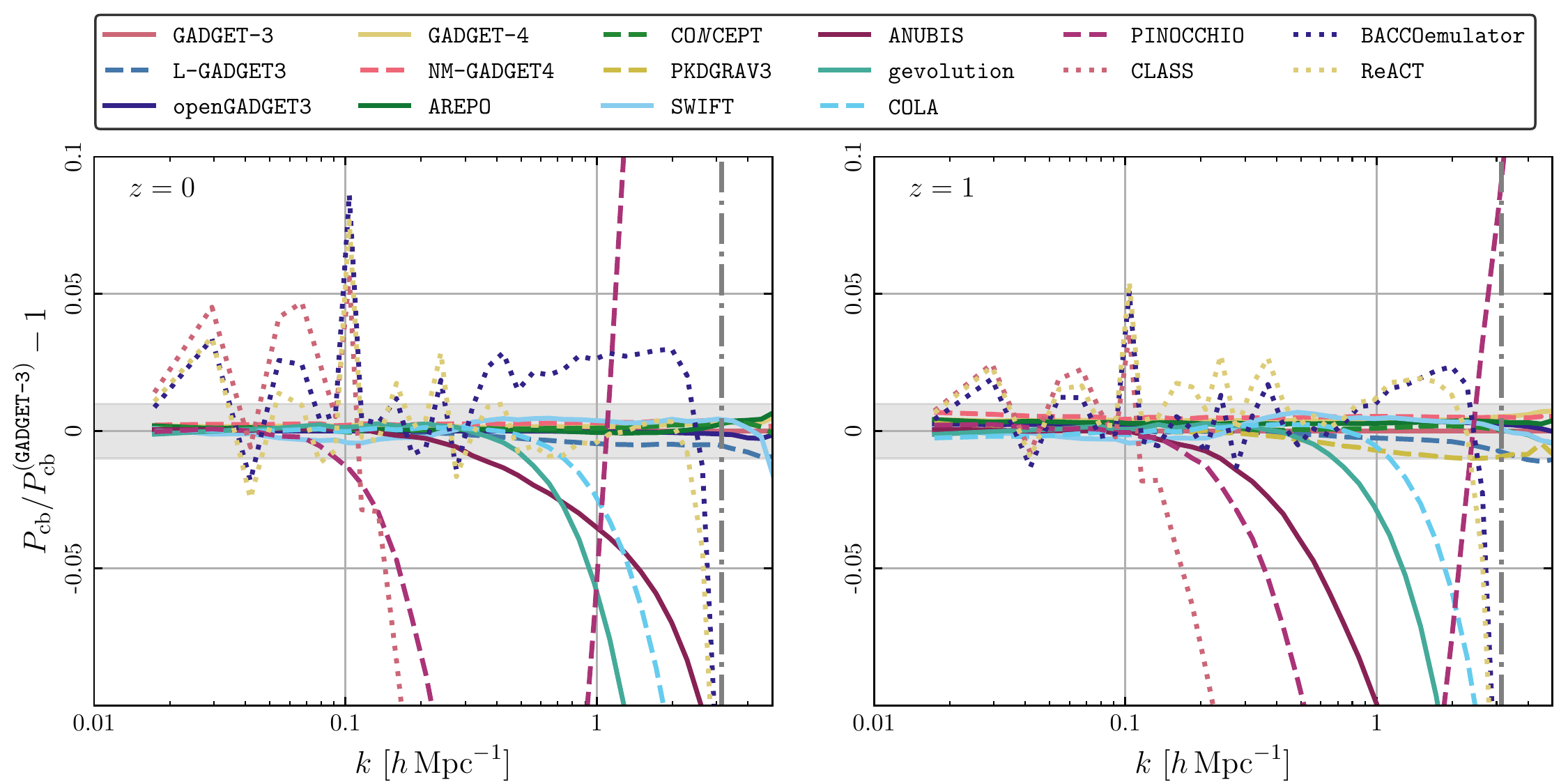}
  \caption{Power spectrum of CDM and baryons as measured from different codes, relative to \GADGET{3} at $z = 0$ and $z = 1$ for a neutrino mass of $\sum m_\nu  = 0.15\rm{eV}$ in the simulation with  $L_\mathrm{box} = 512\,h^{-1}\,\mathrm{Mpc}$ and $N_\mathrm{part} = 512^3$ CDM and baryon particles. The corresponding particle Nyquist wavenumber is indicated by the grey dash-dotted line. The grey bands highlight the interval of $\pm\,0.01$.}
  \label{fig:pk_ratios_fiducial}
\end{figure}

The leading contribution to $P_\text{m}(k)$ is $P_\cb(k)$, which is suppressed in massive neutrino models. Figure~\ref{fig:pk_ratios_fiducial} shows the ratio of $P_\cb(k)$ for models with a neutrino mass of $\sum m_\nu  = 0.15\,\rm{eV}$ relative to the \GADGET{3} simulation which we arbitrarily pick as the reference. In all our figures, results from codes where massive neutrinos are represented through an $N$-body ensemble are plotted using solid lines, while other $N$-body methods, including surrogates, use dashed lines. Any additional predictions use dotted lines. In Figure~\ref{fig:pk_ratios_fiducial}, the linear prediction, computed with \CLASS, is shown by the rose dotted line that drops off sharply at $k \approx 0.1\,h\,\mathrm{Mpc}^{-1}$ beyond which the error quickly exceeds $10\%$. At $z=0$, on the largest scales, all codes deviate less than 1\% --- the fluctuations seen in the dotted lines are largely due to the lack of cosmic variance in the codes that predict $P_\mathrm{cb}(k)$ directly. On smaller scales, some of the codes start to deviate from the \GADGET{3} reference. \PINOCCHIO is in agreement within 1\% up to $k \approx 0.1\,h\,\rm{Mpc}^{-1}$, \ANUBIS up to $k \approx 0.3\,h\,\rm{Mpc}^{-1}$, \gevolution up to $k \approx 0.5\,h\,\rm{Mpc}^{-1}$, and \COLA up to $k \approx 0.7\,h\,\rm{Mpc}^{-1}$. The other codes stay within a 1\% deviation from \GADGET{3} for all scales down to the particle Nyquist scale of $k_\Nyq=\pi\,N_\mathrm{part}^{1/3}\,L_\mathrm{box}^{-1}\approx 3\,h\,\rm{Mpc}^{-1}$. This scale is indicated by a vertical dash-dotted line.
The \ReACT and \BACCOemulator codes stay accurate within 1\% to 5\% on all scales down to the Nyquist scale. At $z=1$ the qualitative behaviour is similar, but most notably the codes disagree more on large scales while still staying within a 1\% agreement. Here, \PINOCCHIO{} is accurate up to $k \approx 0.2\,h\,\rm{Mpc}^{-1}$.

\begin{figure}
  \includegraphics[width=\textwidth]{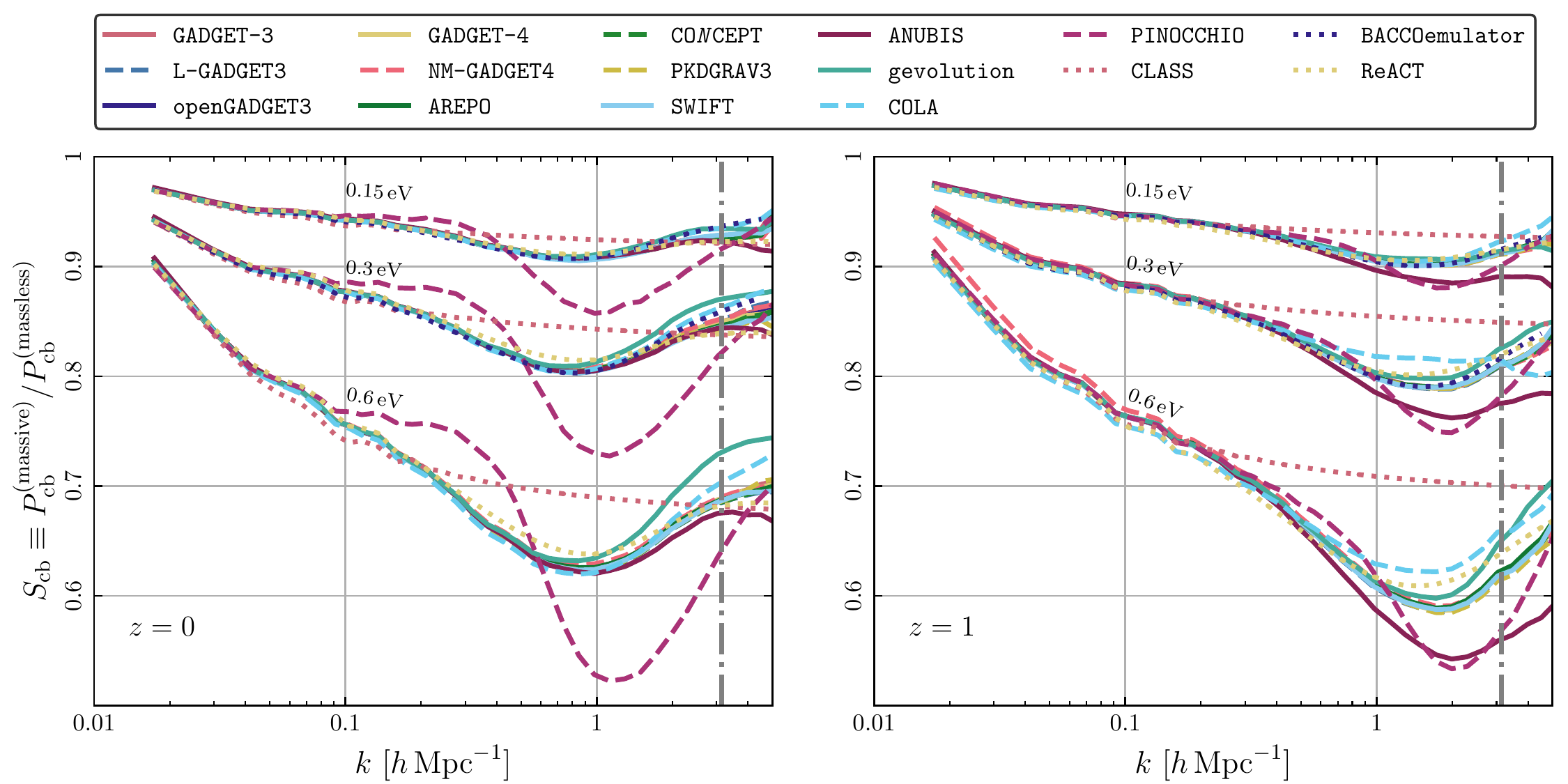}
  \caption{Suppression of the power spectrum of CDM and baryons at $z=0$ and $z=1$ for three different neutrino masses, $\sum m_\nu \in \{0.15,\,0.3,\,0.6\}\,\rm{eV}$, when compared to the massless case. Results are from the simulations with $L_\mathrm{box} = 512\,h^{-1}\,\mathrm{Mpc}$ and $N_\mathrm{part} = 512^3$. The corresponding particle Nyquist wavenumber is indicated by the grey dash-dotted line.}
  \label{fig:pk_suppression_fiducial}
\end{figure}

Figure~\ref{fig:pk_suppression_fiducial} shows the ratio of $P_\cb(k)$ for models with massive neutrinos relative to the massless case for $\sum m_\nu \in \{0.15,\,0.3,\,0.6\}\,\rm{eV}$. For later convenience, we denote this suppression ratio by $S_\cb(k)$, or in general
\begin{equation}
    S_\mathrm{X}(k) = \frac{P_\mathrm{X}^\mathrm{(massive)}(k)}{P_\mathrm{X}^\mathrm{(massless)}(k)}\,,
\end{equation}
where the subscript ``X'' denotes any component in question, here $\mathrm{X} \to \cb$. Linear calculations (taken from \CLASS) predict that, on the largest scales, the growth of structure is mostly unaffected so that $S_\cb(k)$ approaches unity, while on small scales $S_\cb(k)$ reaches a plateau. Compared to this linear expectation, all codes in the comparison predict slightly less suppression around
$k=0.1\,h\,\rm{Mpc}^{-1}$ and a much greater suppression for $k>0.3\,h\,\rm{Mpc}^{-1}$, followed by an upturn on nonlinear scales. This upturn has been repeatedly demonstrated and results from the reduced sensitivity of the one-halo contribution \cite{Massara:2014kba,Hannestad:2020rzl}. At $z=0$, we obtain excellent agreement between all simulations and most approximate methods up to the scale of maximum suppression at $k_{\rm{max}} \approx 1\,h\,\rm{Mpc}^{-1}$, where the suppression is $20\%$ greater than the linear prediction. At $z=1$, the scale of maximum suppression shifts to $k_{\rm{max}} \approx 2\,h\,\rm{Mpc}^{-1}$ and the differences are greater, both with the linear prediction and between the codes, with the exception of \PINOCCHIO\footnote{The reason for this disagreement is due to the fact that the accuracy of LPT-based \PINOCCHIO depends on the level of nonlinearity that varies with neutrino mass.} which fares significantly better compared to $z=0$.

\begin{figure}
  \includegraphics[width=\linewidth]{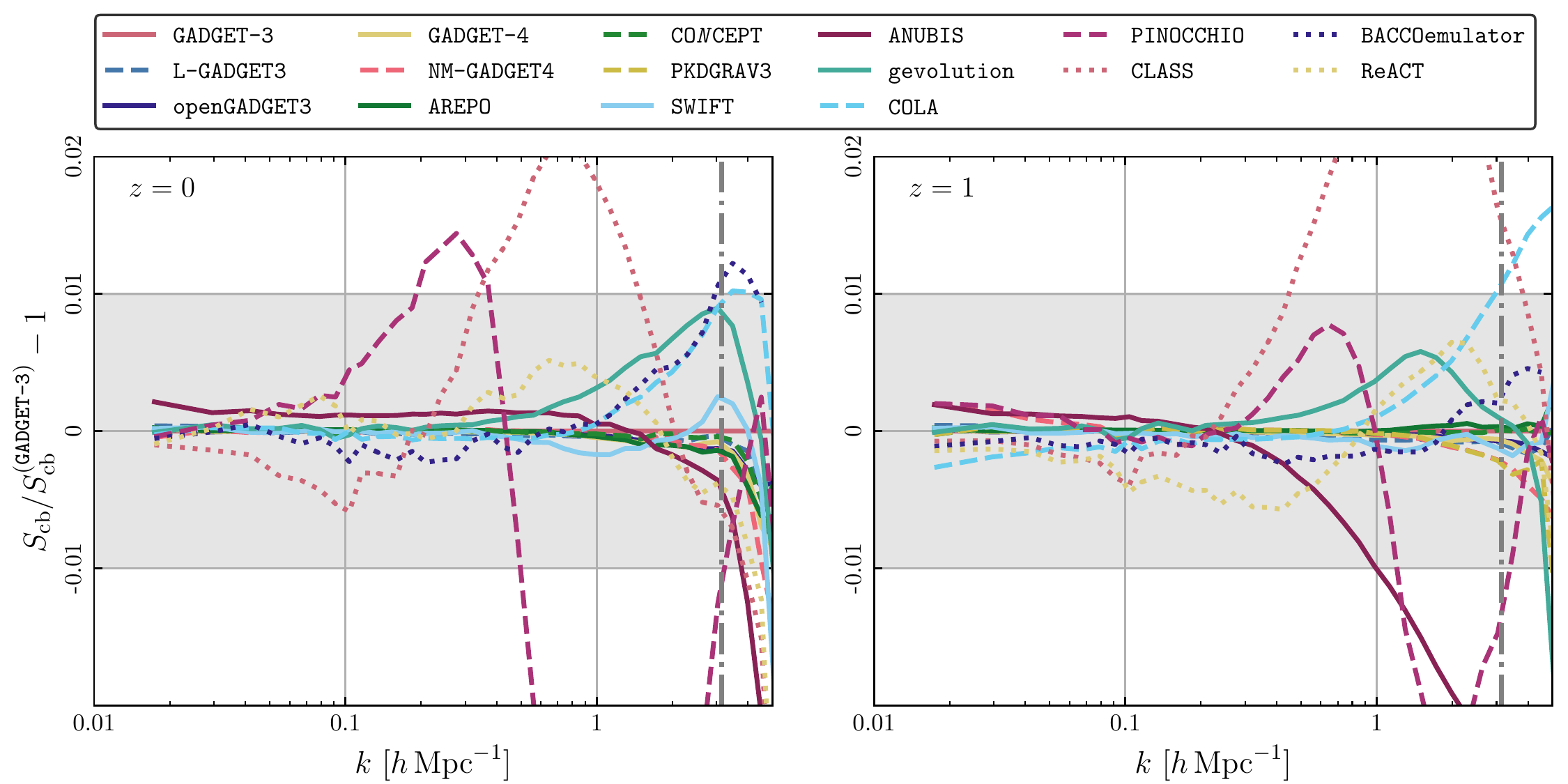}
  \caption{The suppression of the power spectrum of CDM and baryons relative to the one measured in the \GADGET{3} reference runs at $z=0$ and $z=1$ for $\sum m_\nu = 0.15\,\rm{eV}$ for the simulations with $L_\mathrm{box}=512\,h^{-1}\,\mathrm{Mpc}$ and $N_\mathrm{part} = 512^3$. The corresponding particle Nyquist wavenumber is indicated by the grey dash-dotted line. The grey bands highlight the interval of $\pm\,0.01$.}
  \label{fig:pk_ratios_of_ratios_150}
\end{figure}

To study these relative differences in greater detail, we show $S_\cb(k)$ for the smallest neutrino mass, $\sum m_\nu=0.15\,\rm{eV}$, relative to the \GADGET{3} prediction for this quantity in Figure~\ref{fig:pk_ratios_of_ratios_150}. With the exception of \PINOCCHIO and \CLASS, all codes agree to better than $1\%$ at $z=0$ up to the particle Nyquist scale.  Near $k_\Nyq$, the approximate \COLA method and the \BACCOemulator{} differ by more than $1\%$ from the bulk of the simulations, while \gevolution differs by slightly less than 1\%. Beyond this scale, the predictions diverge and should be compared to higher-resolution runs since our estimator of the power spectrum is computed on a mesh with a matching Nyquist scale. The measured power spectra therefore cannot be used beyond $k_\Nyq$ where they become strongly biased. This particularly affects the comparison between simulations --- where the power spectrum estimator is employed --- and other methods to predict $P_\mathrm{cb}(k)$. 
At $z=1$, nonlinearities are smaller and the agreement between the simulations is better. Here the snapshot produced by {\PINOCCHIO} achieves percent accuracy to $k \approx 0.3\,h\,\rm{Mpc}^{-1}$.
Some of the approximate methods fare slightly worse at this earlier time compared to $z=0$, with the difference between \CLASS and the simulations increasing by $50\%$ and \COLA diverging beyond $k=2\,h\,\rm{Mpc}^{-1}$. The \NMGADGET method requires an additional post-processing coordinate transformation at any redshift except $z=0$. Because this additional step is omitted in this work for simplicity, a small error at low wavenumbers remains in the $z=1$ data. This explains why the error is larger at that redshift than at the final time.

\ANUBIS{} notably drops off for $k>1\,h\,\rm{Mpc}^{-1}$ at $z=1$ and also shows a small excess on linear scales both at $z=0$ and $z=1$. This excess originates from the massless case and is a result of \ANUBIS{} being run with a coarser base grid than the other codes ($512^3$ for all simulations) due to limited resources. A finer base grid requires more memory but the lack of it can be somewhat compensated by using a smaller time step. For the \ANUBIS{} massive neutrino runs, this is done automatically but for the massless case the time step has been set to half of that originally calculated by the code. Tests indicate that further reducing the time step or ideally using a finer base grid should lessen the excess at large scales, but finding the optimal choice of code settings is not the aim of this work. The drop-off observed for \ANUBIS{} at $z=1$ for $k>1\,h\,\rm{Mpc}^{-1}$ is due to differences in resolution between the various simulations. As \ANUBIS{} is an AMR-code, a modified version of the \RAMSES code originally written by Teyssier~\cite{Teyssier:2001cp}, the inclusion of massive neutrinos, which suppresses clustering on small scales, also reduces the number of refinements reached in the simulation compared to the massless case.
This effect increases with the neutrino mass and is more noticeable for higher redshifts where there is also less refinement due to less clustering. This can be solved by a higher particle density which automatically leads to more refinement. Generally this is also necessary to find a better agreement between \ANUBIS and \GADGET{3} for smaller scales, as can be observed from the high-resolution runs and also in a previous comparison conducted by Schneider et al.~\cite{Schneider:2015yka} between \RAMSES, \GADGET{3}, and \PKDGRAV.

\begin{figure}
  \includegraphics[width=\textwidth]{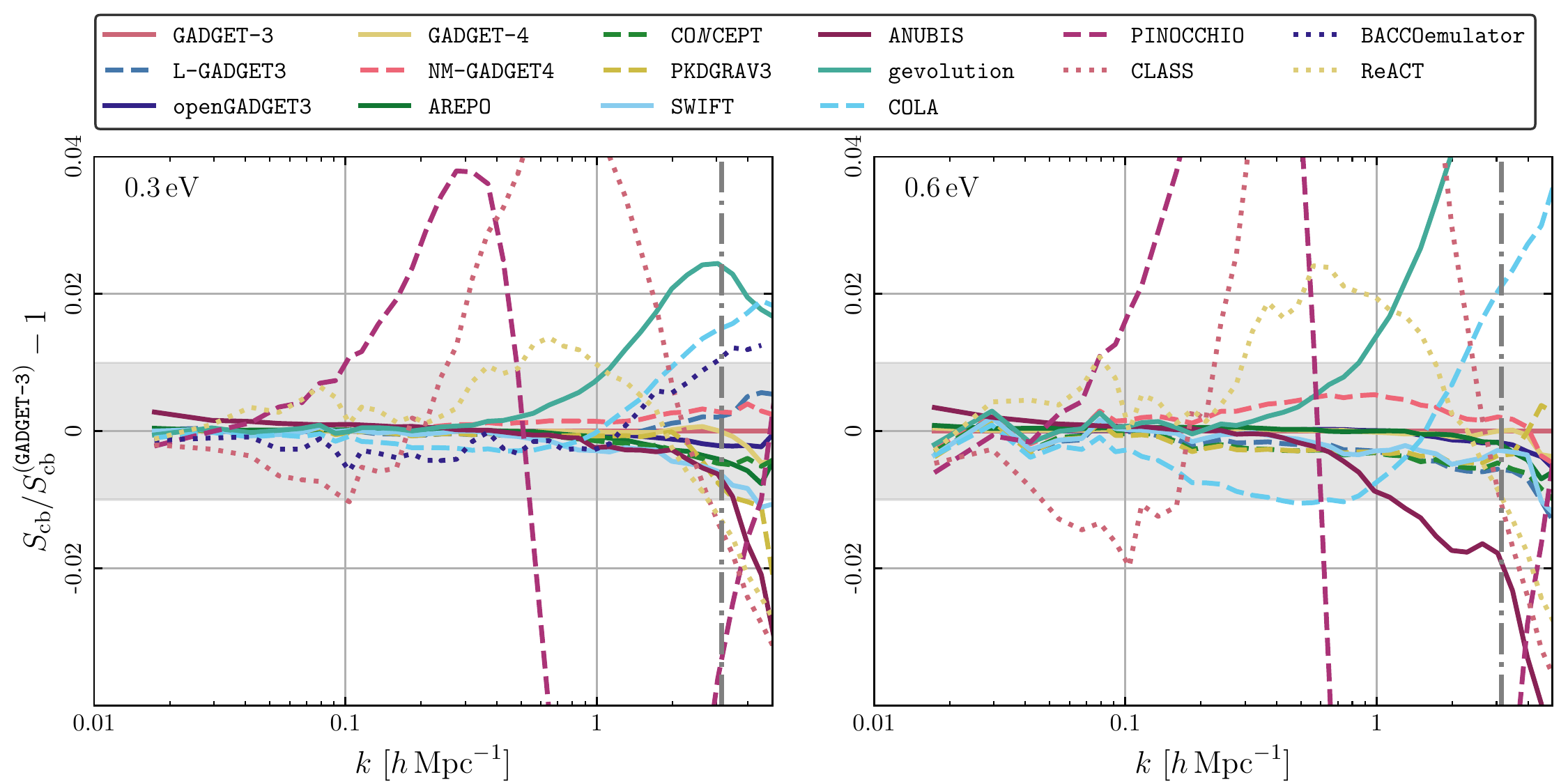}
  \caption{The suppression of the power spectrum of CDM and baryons relative to the one measured in the \GADGET{3} reference runs at $z=0$ for $\sum m_\nu \in \{0.3,0.6\}\,\rm{eV}$ for the simulations with $L_\mathrm{box}=512\,h^{-1}\,\mathrm{Mpc}$ and $N_\mathrm{part} = 512^3$. The corresponding particle Nyquist wavenumber is indicated by the grey dash-dotted line. The grey bands highlight the interval of $\pm\,0.01$.}
  \label{fig:pk_ratios_of_ratios_300_600}
\end{figure}

The observed differences are greater when the neutrino mass is increased, especially for the approximate methods. Figure~\ref{fig:pk_ratios_of_ratios_300_600} shows $S_\cb(k)$ relative to the \GADGET{3} prediction for models with $\sum m_\nu = 0.3\,\rm{eV}$ (left panel) and $\sum m_\nu = 0.6\,\rm{eV}$ (right panel) at $z=0$. While the \BACCOemulator agrees to better than $1\%$ with \GADGET{3} up to $k_\Nyq$ for $0.3\,\rm{eV}$, it makes no prediction for $0.6\,\rm{eV}$ because that value lies far outside the range covered by the training set of the emulator. We deliberately include a case of such a large neutrino mass to exacerbate the differences between the various numerical implementations. The reaction method differs by slightly more than $1\%$ and $2\%$ at $k=0.7\,h\,\rm{Mpc}^{-1}$ for total neutrino masses of $0.3\,\rm{eV}$ and $0.6\,\rm{eV}$, respectively. The differences with the linear prediction (\CLASS) and with \PINOCCHIO{} are large, as noted above for $0.15\,\rm{eV}$.

The agreement is excellent for the other codes, but some subtle differences can be discerned. On large scales, we observe that \SWIFT, \CONCEPT, \COLA, \gevolution, \PKDGRAV, \NMGADGET, and \LGADGET show the same coherent scatter relative to \GADGET{3}, especially for a neutrino mass of $0.6\,\rm{eV}$. This is due to the contamination from shot noise in the neutrino particle implementation used by the \GADGET{3} run. The other mentioned codes have implementations that do not suffer from shot noise or take measures to limit the contamination. For instance, in \gevolution the neutrino $N$-body ensemble is evolved throughout the simulation, but it is only used as source of gravitational fields from redshift $z=7$ and below. At higher redshifts, the code uses the linear grid-based density instead. The reasoning behind this strategy is that shot noise is constant over time and hence more problematic at high redshift where cosmological perturbations are smaller in comparison. On the other hand, the linear prediction for neutrinos is expected to be very accurate at high redshift. One can therefore reduce the total error by judiciously choosing the time at which the code switches from linear to fully nonlinear neutrino treatment. It is nonetheless reassuring that even without mitigating against shot noise the scatter remains far below $1\%$.

On small scales the differences are also larger. The lines for \COLA{} and \gevolution{} track each other closely, but differ by more than $1\%$ from the other codes for $k>1\,h\,\rm{Mpc}^{-1}$. \SWIFT{}, \PKDGRAV, and \CONCEPT{} are low compared to \GADGET{3} for $k>1\,h\,\rm{Mpc}^{-1}$, unlike what was seen in Figure~\ref{fig:pk_ratios_of_ratios_150}. \ANUBIS{} diverges from \GADGET{3} by more than $1\%$ for $k>1\,h\,\rm{Mpc}^{-1}$ for the $0.6\,\rm{eV}$ neutrino mass case. As mentioned earlier, this is due to the 
fact that the AMR scheme has a lower effective resolution as the neutrino mass increases, simply because refinement is triggered by clustering. Finally, \NMGADGET{} is slightly higher than \GADGET{3}. However, these differences remain below $1\%$ well beyond $k_\Nyq$.

\subsubsection{Convergence tests}
\label{subsec:convergencePS}

\begin{figure}
  \includegraphics[width=\textwidth]{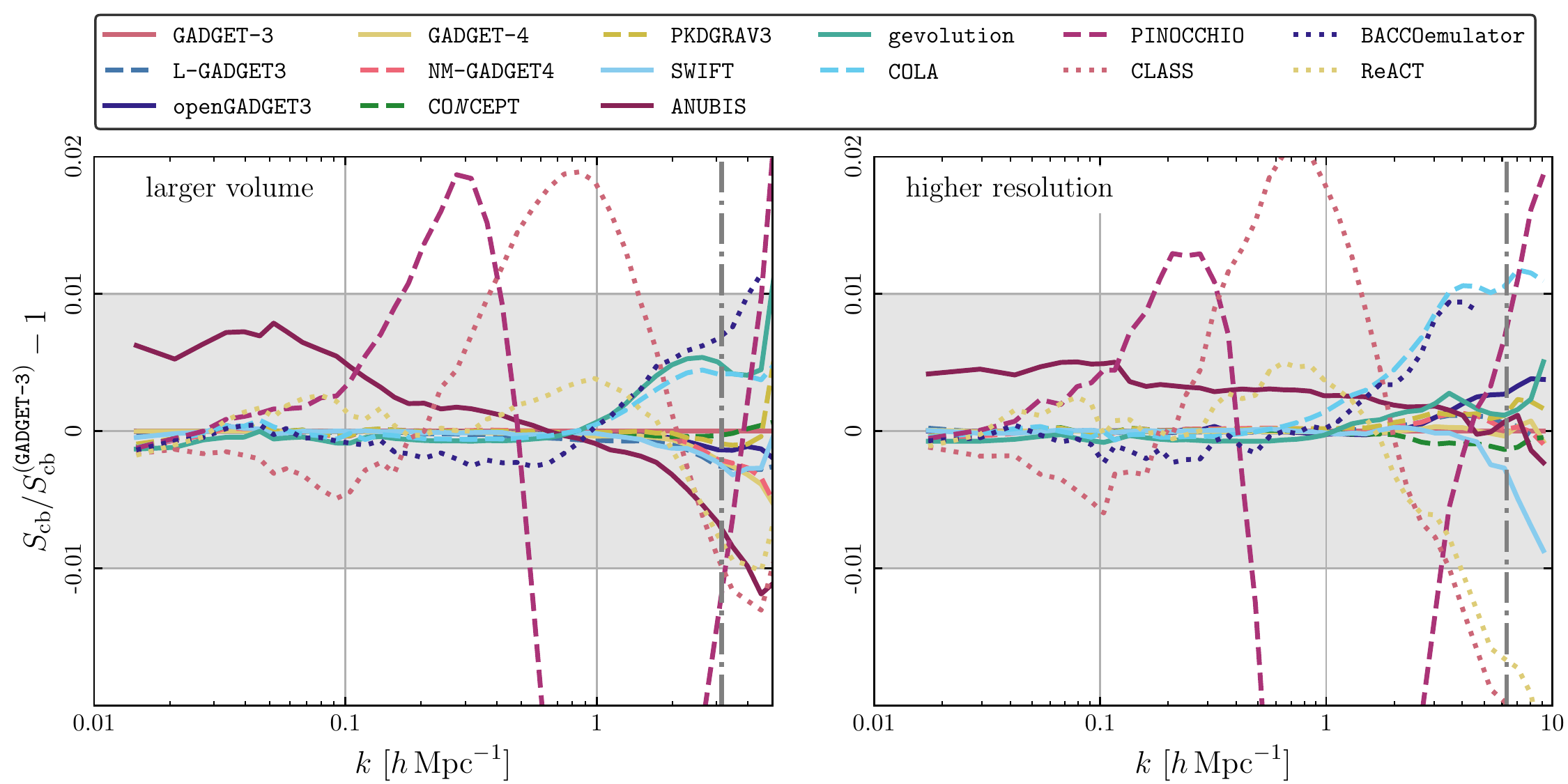}
  \caption{The suppression of the power spectrum of CDM and baryons relative to the one measured in the \GADGET{3} reference runs at $z=0$ for $\sum m_\nu = 0.15\,\rm{eV}$ for the simulations with a larger volume, $L_\mathrm{box} = 1024\,h^{-1}\,\mathrm{Mpc}$ and $N_\mathrm{part} = 1024^3$ (left panel), and at a higher resolution in the small volume, $L_\mathrm{box} = 512\,h^{-1}\,\mathrm{Mpc}$ and $N_\mathrm{part} = 1024^3$ (right panel). The respective particle Nyquist wavenumbers are indicated by the grey dash-dotted lines. The grey bands highlight the interval of $\pm\,0.01$.}
  \label{fig:pk_ratios_of_ratios_150_res_and_vol}
\end{figure}

To study the numerical convergence of our results, we consider the effects of finite box size and resolution. Figure~\ref{fig:pk_ratios_of_ratios_150_res_and_vol} shows the relative suppression for the runs with larger volume (left panel) and higher resolution (right panel). When $L_\mathrm{box}$ is doubled at fixed resolution, the agreement remains excellent on linear scales and is sometimes even slightly better around $k_\Nyq$, providing an important consistency check for most codes. Increasing instead the mass resolution by doubling $k_\Nyq$, the agreement between the simulations improves significantly on nonlinear scales. Including more scales in either direction, most codes remain within $1\%$ of the reference runs done with \GADGET{3}. The excess on large scales for the case of \ANUBIS persists as a result of the coarse base grid. For the runs with $N_\mathrm{part} = 1024^3$, this base grid is even less suited and the time-steppings for the massless cases are set to $0.1$ and $0.15$ times the original time step for the larger-box and high-resolution runs, respectively.

\subsubsection{Contributions from neutrinos}

\begin{figure}
  \includegraphics[width=\textwidth]{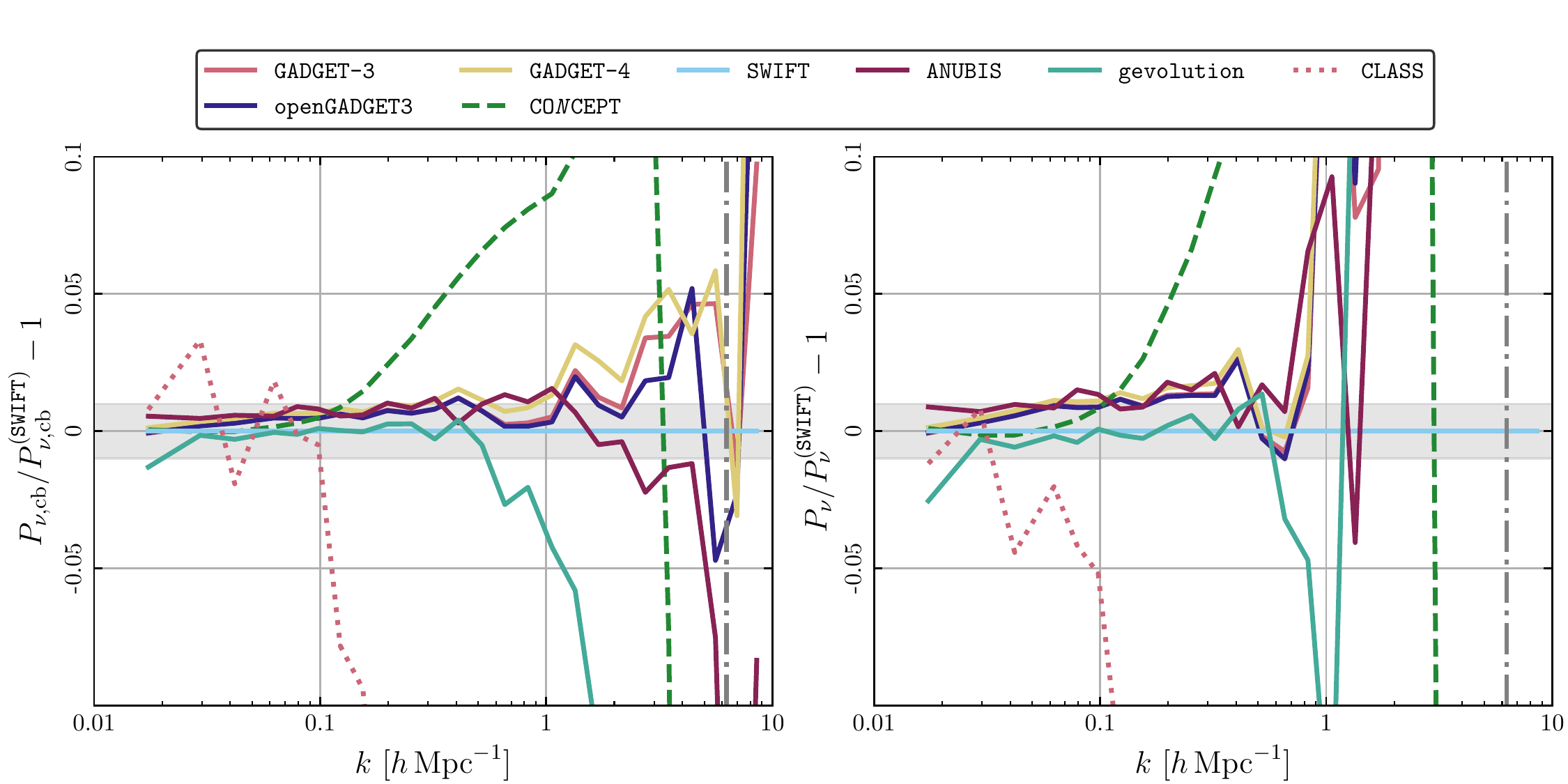}
  \caption{Relative cross-power spectrum of neutrinos with CDM and baryons (left) and neutrino auto-power spectrum (right) with respect to \SWIFT{} at $z=0$ for $\sum m_\nu = 0.15\,\rm{eV}$ for the higher-resolution simulation with $L_\mathrm{box}=512\,h^{-1}\,\mathrm{Mpc}$ and $N_\mathrm{part} = 1024^3$.
  The Nyquist wavenumber is indicated by the grey dash-dotted line. The grey bands highlight the interval of $\pm\,0.01$.}
  \label{fig:xpk_pk_neutrino_150}
\end{figure}

The subdominant contributions to $P_\text{m}(k)$ are the cross-power spectrum $P_{\nu,\cb}(k)$ between neutrinos and the CDM and baryon component, and the auto-power spectrum $P_{\nu}(k)$ of neutrinos. As can be seen from Eq.~\eqref{eq:Pm_contributions}, these are suppressed by the small factors $f_\nu$ and $f_\nu^2$, respectively, and they are themselves additionally strongly suppressed with respect to $P_\cb(k)$ on scales smaller than the neutrino free-streaming scale. While both of these contributions will be exceedingly hard to constrain individually from observations, it is nonetheless interesting to study them in the context of our code comparison in order to highlight some more subtle differences in the numerical schemes.
Figure~\ref{fig:xpk_pk_neutrino_150} (left panel) shows $P_{\nu,\cb}(k)$ for various codes relative to the result from the \SWIFT code, for the smallest neutrino mass, $\sum m_\nu=0.15\,\rm{eV}$, computed from the high-resolution simulations. We use \SWIFT as the reference here because it has a very low level of shot noise in the neutrino component, yet is able to track the nonlinear evolution of neutrinos very accurately. As is the case with CDM and baryons, linear theory cannot describe neutrino clustering on nonlinear scales and therefore \CLASS significantly underestimates the cross-power spectrum for $k>0.1\,h\,\rm{Mpc}^{-1}$. 
Most other codes use a particle implementation of neutrinos and scatter about the \SWIFT prediction partially due to shot noise.
The results from \CONCEPT have no shot noise, but depart from the other codes for $k>0.2\,h\,\rm{Mpc}^{-1}$. The relative difference to \SWIFT is $8\%$ at $k=1\,h\,\rm{Mpc}^{-1}$. Much the same applies to the neutrino auto-power spectrum, $P_\nu(k)$, 
shown relative to the neutrino auto-power spectrum of \SWIFT in the right panel of 
Figure~\ref{fig:xpk_pk_neutrino_150}. Here, for the codes with a neutrino particle ensemble, the dominant contribution to the shot noise is removed. Except for \SWIFT this is done by subtracting the inverse of the average neutrino particle density, $\bar{n}^{-1}$, from the measured power spectrum, where $\bar{n}^{-1} = 0.125\,h^{-3}\,\mathrm{Mpc}^{3}$ in our high-resolution simulations. For \SWIFT the subtracted values are derived from the $\delta f$ method. The difference between \SWIFT and \CONCEPT is $19\%$ (off the chart) at $k=1\,h\,\rm{Mpc}^{-1}$ and the effects of shot noise are even more evident in the other codes. The agreement between \gevolution and \SWIFT{} is quite remarkable, and on the largest scales these results are also more consistent with \CONCEPT than with the other $N$-body codes.

\subsubsection{Total matter}

\begin{figure}
  \includegraphics[width=\textwidth]{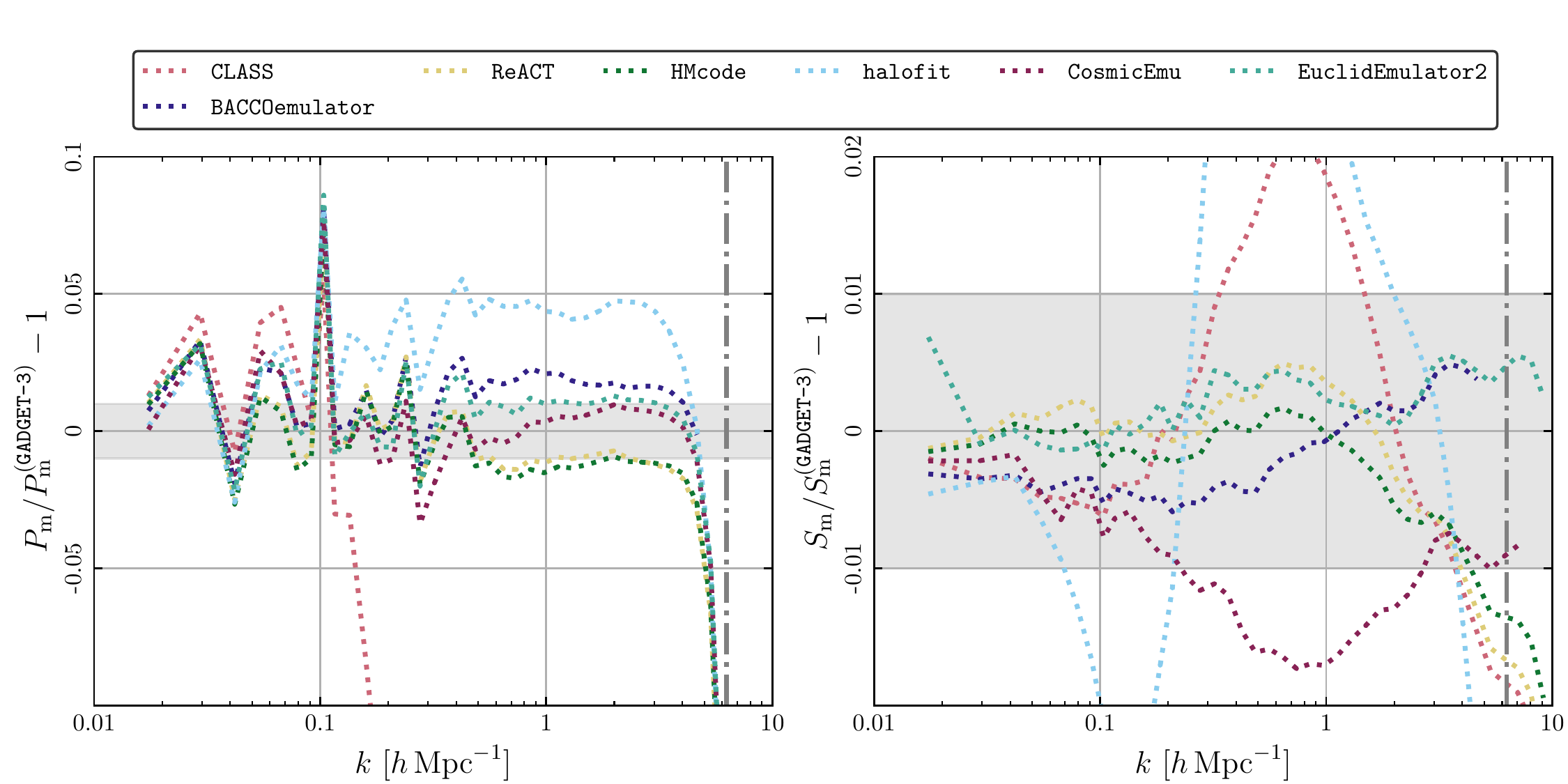}
  \caption{Total matter power spectrum $P_\text{m}(k)$ at $z=0$ for $\sum m_\nu=0.15\,\rm{eV}$ from emulators and fitting methods (left panel), and the respective suppression $S_\text{m}(k)$ with respect to the massless case (right panel), compared to the higher-resolution reference \GADGET{3} simulation with $L_\text{box}=512\,h^{-1}\,\rm{Mpc}$ and $N_\mathrm{part} = 1024^3$. The corresponding particle Nyquist wavenumber of the \GADGET{3} run is indicated by the grey dash-dotted line and marks the limit beyond which the estimator of the power spectrum becomes unreliable. Disagreements near and beyond this line are therefore not indicative of errors in the emulators. The grey bands highlight the interval of $\pm\,0.01$.}
  \label{fig:emulator_pk_neutrino_150}
\end{figure}

Some of the codes, in particular the \EuclidEmulator, only provide predictions for the power spectrum of total matter, $P_\mathrm{m}(k)$. In Figure~\ref{fig:emulator_pk_neutrino_150}, we show the relative agreement of different emulators and other codes predicting $P_\mathrm{m}(k)$ to our high-resolution reference run with \GADGET{3}. The left panel shows the result for the matter power spectrum itself, $P_\mathrm{m}(k)$, at $\sum m_\nu = 0.15$\,eV. The correlated fluctuations on large scales are due to sample variance which is only present in the reference simulation and not in the predicted spectra. We note that emulators, \HMcode, and the halo-model reaction method perform slightly better than the fitting recipe of \halofit. The results of the \EuclidEmulator are marginally consistent with the claimed accuracy of $1\%$, but the neutrino mass lies close to the boundary of parameter space the emulator was trained for. The right panel of Figure~\ref{fig:emulator_pk_neutrino_150} shows corresponding results for the power suppression factor, $S_\mathrm{m}(k)$, with respect to the massless scenario. \CosmicEmu shows a disagreement larger than $1\%$ around $k \approx 1\,h\,\mathrm{Mpc}^{-1}$ where the power suppression is largest. Also \halofit performs poorly, worse even than linear theory (\CLASS). The other codes remain within $1\%$ of the \GADGET{3} result up to the particle Nyquist scale. Overall our results are in fair agreement with the detailed comparison carried out by Parimbelli et al.~\cite{Parimbelli:2022pmr}.

\subsection{Bispectra}

The nonlinear evolution of matter fluctuations generates a non-vanishing bispectrum, the three-point correlation function of matter in Fourier space, even if non-Gaussianity is negligibly small in the initial conditions. This represents an opportunity for the measurement of neutrino masses, as we expect that the suppression predicted by linear theory is enhanced at the nonlinear level \cite{Saito:2008bp,Wong:2008ws,Audren:2012vy,deBelsunce:2018xtd}. The total matter bispectrum in the presence of massive neutrinos can be schematically defined as
\begin{multline}
    B_\mathrm{mmm}(k_1,k_2,k_3) = (1-f_\nu)^3\, B_\mathrm{ccc}(k_1,k_2,k_3)+f_\nu\,(1-f_\nu)^2\, B^{(\mathrm{sym})}_{\mathrm{cc}\nu}(k_1,k_2,k_3)\\
    +f_\nu^2\,(1-f_\nu)\,B^{(\mathrm{sym})}_{\mathrm{c}\nu\nu}(k_1,k_2,k_3) + f_\nu^3\, B_{\nu\nu\nu}(k_1,k_2,k_3),
\end{multline}
where ``ccc'' denotes the CDM and baryon bispectrum (we do not write ``cb\,cb\,cb'' to avoid clutter) and we note the presence of cross cold-cold-neutrino ``$\mathrm{cc}\nu$'' and cold-neutrino-neutrino ``$\mathrm{c}\nu\nu$'' terms that are symmetrised, as indicated by the superscript ``$(\mathrm{sym})$''. In this work, we focus on the bispectrum of CDM and baryons only, but cross terms have also been investigated in the literature, e.g.\ by Ruggeri et al.~\cite{Ruggeri:2017dda}. As for the power spectrum case, the leading term is the one of CDM and baryons. 

\begin{figure}
    \centering
    \includegraphics[width=\textwidth]{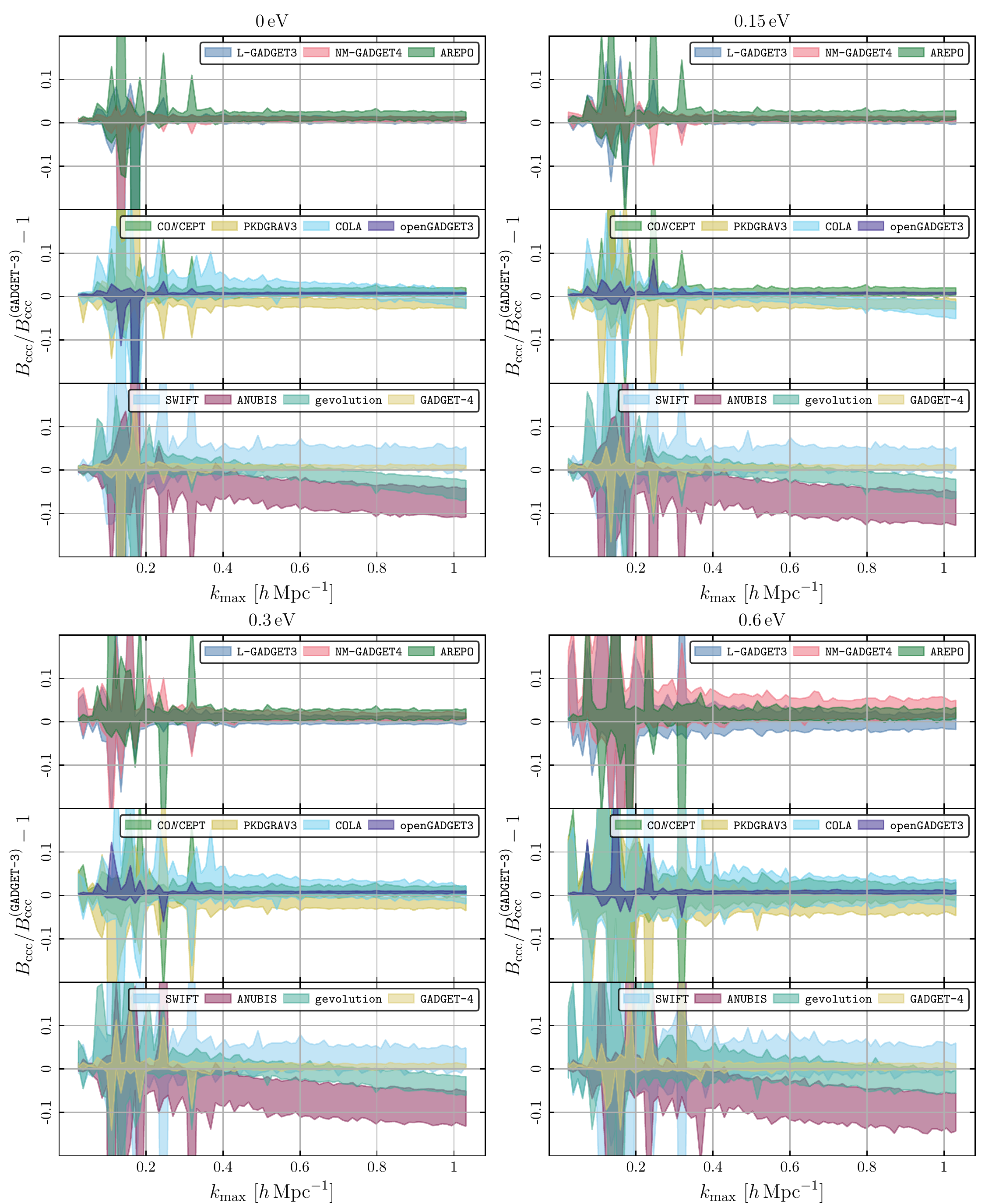}
  \caption{The coloured bands show the scatter of $B_\mathrm{ccc}$ for all triangle configurations with respect to the measurement from \GADGET{3} at redshift $z = 1$. The horizontal axis indicates the maximum wavenumber in each triangle configuration, $k_\mathrm{max} = \max(k_1, k_2, k_3)$, in order to present the results in a simple plot. The four panels show different neutrino masses. Each plot uses three subpanels to make the results of the many different codes more visible.} 
  \label{fig:allbis}
\end{figure}

\begin{figure}
    \centering
    \includegraphics[width=\textwidth]{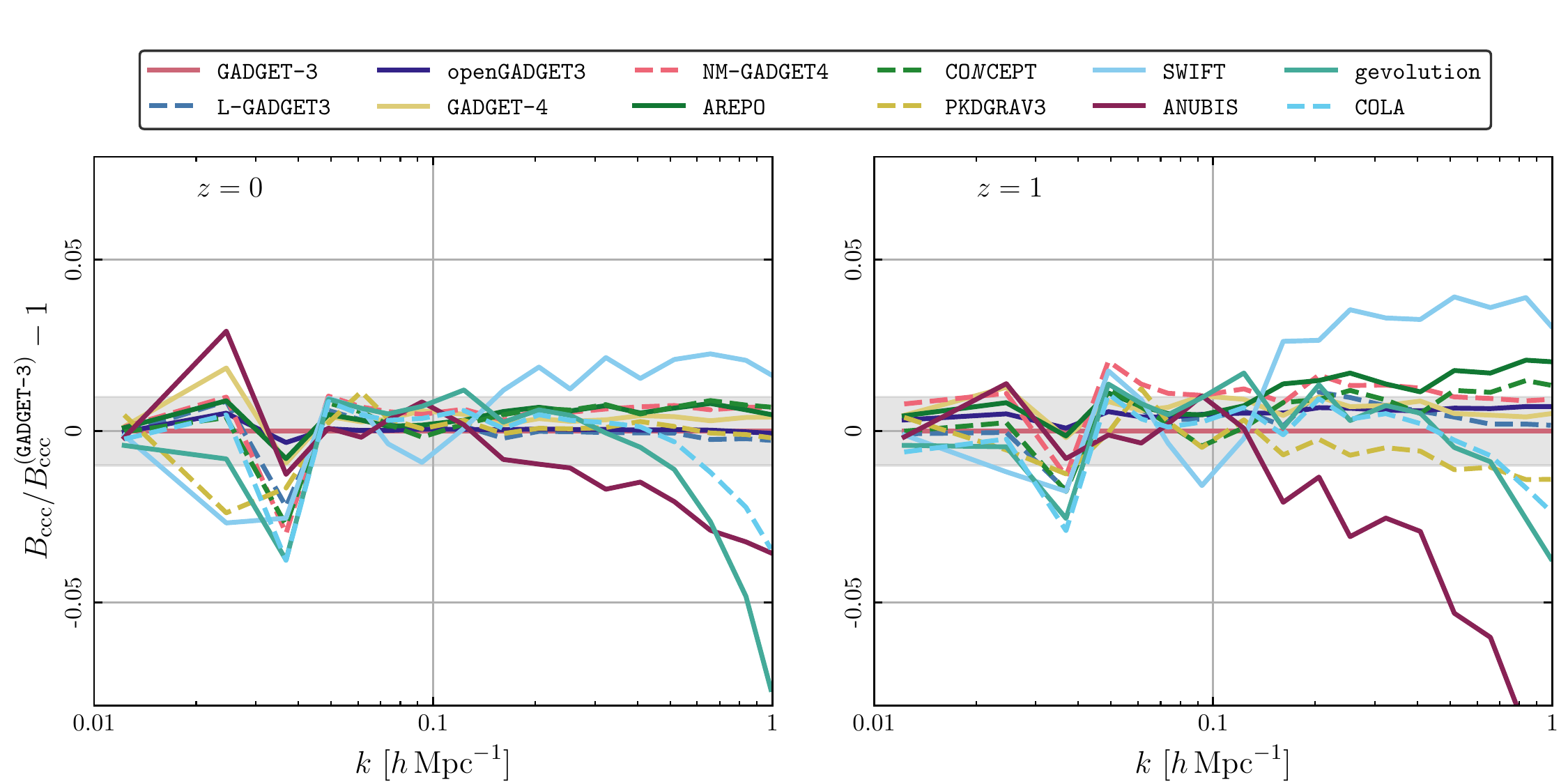}
   \caption{Squeezed bispectrum of CDM and baryons as measured from different codes, relative to \GADGET{3} at $z = 0$ and $z = 1$ for a neutrino mass of $\sum m_\nu  = 0.15\rm{eV}$ in the simulation with  $L_\mathrm{box} = 512\,h^{-1}\,\rm{Mpc}$ and $N_\mathrm{part} = 512^3$ CDM and baryon particles. Results shown are for the squeezed configuration where $k_1 = k_2 \equiv k$, $k_3 = 0.012\,h\,\mathrm{Mpc}^{-1}$. The grey bands highlight the interval of $\pm\,0.01$.}
  \label{fig:squeezedvsGADGET-3}
\end{figure}

Figure \ref{fig:allbis} shows a comparison of the bispectrum $B_\mathrm{ccc}$ as measured at redshift $z=1$ by all simulation codes\footnote{\PINOCCHIO is not used in this test that is mostly focused on nonlinear scales beyond its range of validity.}
for all triangle configurations and all neutrino masses considered, for the set of runs with $N_\mathrm{part} = 512^3$. Triangles are plotted as a function of $k_\mathrm{max} = \max(k_1, k_2, k_3)$. The comparison shows the maximum percentage difference with respect to \GADGET{3} of all triangle configurations at each value of $k_\mathrm{max}$. For \LGADGET, \openGADGET, \GADGET{4}, \PKDGRAV, \CONCEPT, and \NMGADGET, discrepancies are mostly within $5\%$ for $\sum m_\nu = 0.0$ eV, while steadily growing up to around $10\%$ for massive neutrino cosmologies. Other simulation codes are within $10\%$ already at $\sum m_\nu = 0.0$ eV. \ANUBIS and \gevolution show some of the strongest deviations at large $k_\mathrm{max}$, but this is mainly a result of finite resolution as we find much better agreement in the higher-resolution runs. At low $k_\mathrm{max}$, strong fluctuations can be observed where measurements can cross zero because of sampling variance, which in turn leads to numerical issues when taking ratios.

In Figure~\ref{fig:squeezedvsGADGET-3}, we consider specifically a squeezed configuration for which $k_1 = k_2$ and $k_3 = 0.012\,h\,\mathrm{Mpc}^{-1}$ and show the agreement between different codes for a total neutrino mass of $\sum m_\nu = 0.15$\,eV at redshift $z=0$ (left panel) and $z=1$ (right panel). We use measurements from our simulations with $L_\mathrm{box} = 512\,\si{\hMpc}$ and $N_\mathrm{part} = 512^3$. The relative agreement is better at low redshift, partially due to the fact that the signal amplitude is larger there.

\begin{figure*}[pt]
    \centering
    \includegraphics[width=\textwidth]{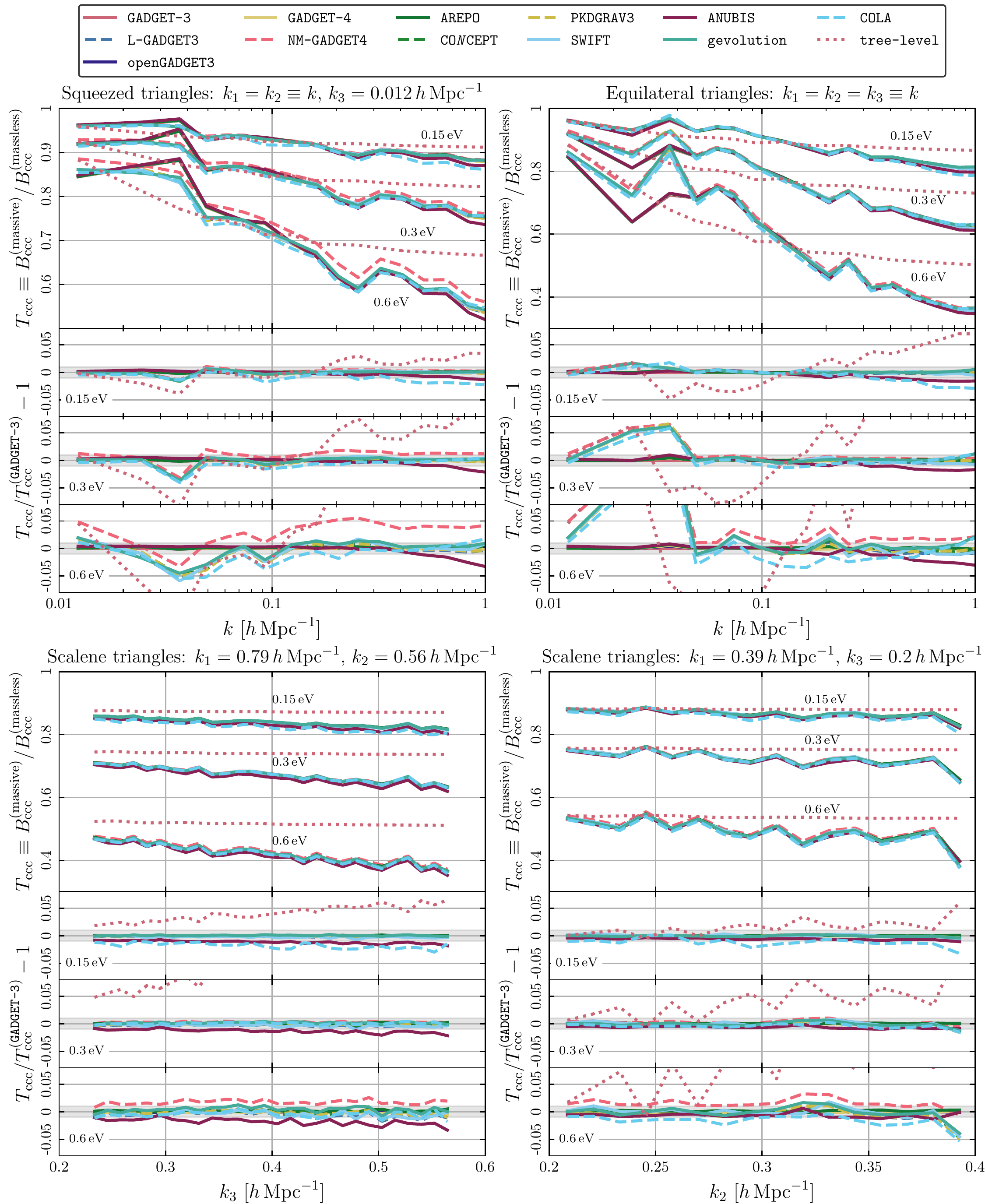}
   \vspace{-4mm}
   \caption{The four panels show bispectrum measurements at redshift $z = 1$ in the simulations with $L_\mathrm{box} = 512\,h^{-1}\,\rm{Mpc}$ and $N_\mathrm{part} = 512^3$ for different triangle configurations: squeezed (top left panel), equilateral (top right panel), and scalene configurations A and B (bottom panels). In each panel, the top subpanel shows the suppression ratio of the bispectrum of CDM and baryons for three different neutrino masses $\sum m_\nu \in \{0.15, 0.3, 0.6\}$\,eV with respect to the massless case, and the three bottom subpanels show the respective relative differences of the various codes when compared to \GADGET{3}.\vspace{-4mm}}
  \label{fig:shapesNB}
\end{figure*}

Figure~\ref{fig:shapesNB} is a comparison of simulation codes for different triangle configurations, for all choices of neutrino masses at redshift $z=1$. We consider four triangle configuations: squeezed, equilateral and two different scalene configurations. Squeezed configurations refer to triangles where one of the side is much shorter than the other two (in Fourier space) which corresponds to looking at the correlation of a distant point with two points close to each other. Equilateral configurations, instead, refer to the correlation of three points at equal distance. Scalene triangles do not have any specific symmetry. In detail, we consider 
\begin{itemize}
    \item {\bfseries squeezed configurations}, for which $k_1 = k_2$ and $k_3 = 0.012\,h\,\mathrm{Mpc}^{-1}$, plotted as a function of $k_1$ as in Figure~\ref{fig:squeezedvsGADGET-3};
    \item {\bfseries equilateral configurations}, for which $k_1 = k_2 = k_3$, plotted as a function of $k_1$;
    \item {\bfseries scalene configurations A}, for which $k_1 =0.79\,h\,\mathrm{Mpc}^{-1}$, $k_2 = 0.56\,h\,\mathrm{Mpc}^{-1}$, plotted as a function of $k_3$;
    \item {\bfseries scalene configurations B}, for which $k_1 =0.39\,h\,\mathrm{Mpc}^{-1}$, $k_3 = 0.2\,h\,\mathrm{Mpc}^{-1}$, plotted as a function of $k_2$.
\end{itemize}
In each of the four panels of Figure~\ref{fig:shapesNB}, the top subpanel shows the ratio between the CDM and baryon bispectrum for massive neutrino cosmologies over massless ones, which we define in analogy to the case of the power spectrum as
\begin{equation}
     T_\mathrm{ccc}(k_1,k_2,k_3) = \frac{B_\mathrm{ccc}^\mathrm{(massive)}(k_1,k_2,k_3)}{B_\mathrm{ccc}^\mathrm{(massless)}(k_1,k_2,k_3)}\,.
\end{equation}
The three bottom subpanels show the relative differences of the measurements of the suppression ratio in the various codes with respect to \GADGET{3} at each of the three neutrino masses, i.e.\ $\sum m_\nu \in \{0.15, 0.3, 0.6\}$\,eV (from top to bottom). For all configurations considered, discrepancies fall broadly within the $5\%$ range. As expected, massive neutrinos suppress the bispectrum of CDM and baryons at all scales, with a stronger effect at smaller scales. For comparison, we also show the tree-level prediction from perturbation theory, using
\begin{equation}\label{eq:btree}
   B_\mathrm{ccc}^\text{(tree-level)}(k_1,k_2,k_3) = 2 F_2(\vk_1, \vk_2) P_\mathrm{cb}^\mathrm{L}(k_1)P_\mathrm{cb}^\mathrm{L}(k_2) + 2\,\textrm{permutations}\,, 
\end{equation}
where
\begin{equation}
    F_2(\vk_1, \vk_2) = \frac 57 + \frac{1}{2} \frac{\vk_1\cdot\vk_2}{k_1 k_2}\left(\frac{k_1}{k_2}+\frac{k_2}{k_1}\right) + \frac 27\frac{(\vk_1\cdot\vk_2)^2}{k_1^2k_2^2}\,,
\end{equation}
and $P_\mathrm{cb}^\mathrm{L}$ is the linear power spectrum of CDM and baryons generated by \CLASS. We can see in Figure~\ref{fig:shapesNB} that the suppression ratio is in good agreement with this prediction when all the three scales $k_1, k_2, k_3$ have a small wavenumber, and that the measured suppression is generally stronger if some of the wavenumbers are large. This is in line with the results seen in the power spectrum, where strong nonlinearities lead to additional suppression.

At the largest scales, i.e.\ when $k_1, k_2, k_3 \lesssim 0.1\,h\,\mathrm{Mpc}^{-1}$, a distinct feature can be discerned which is particularly prominent in the equilateral configuration and for larger neutrino masses: the measurements from the various codes separate into two groups, \GADGET{3}, \openGADGET, \GADGET{4}, \AREPO, and \ANUBIS on the one side, and \LGADGET, \NMGADGET, \CONCEPT, \PKDGRAV, \SWIFT, \gevolution, and \COLA on the other side. The latter group includes all the codes that employ a mesh-based method, and all of them have means to mitigate against shot noise. We therefore suspect that this dichotomy originates from shot noise in the particle method. This could be tested, e.g.\ by increasing the number of neutrino particles until convergence is achieved.

\begin{figure}
    \centering
    \includegraphics[width=\textwidth]{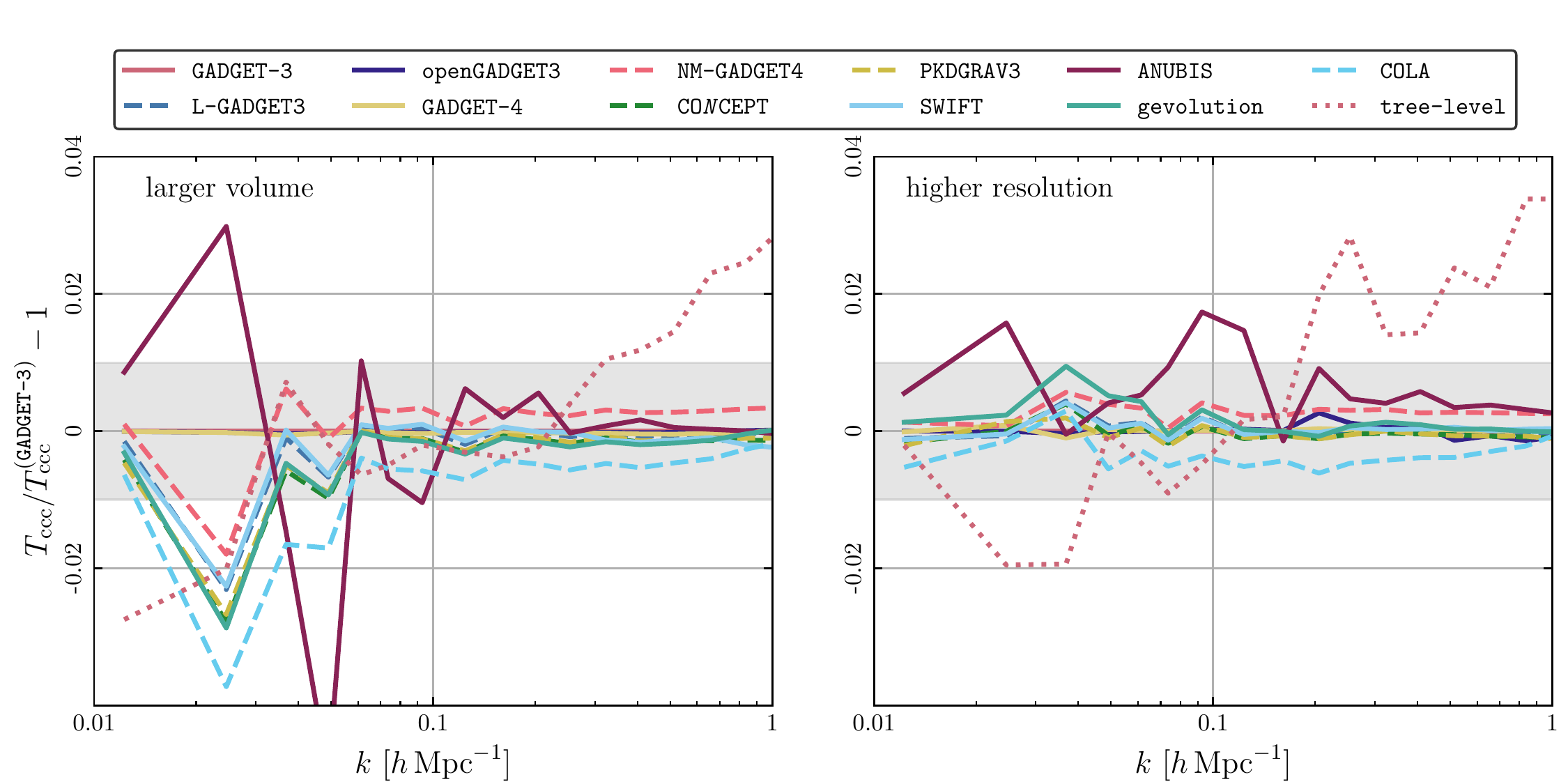}
   \caption{The suppression of the squeezed bispectrum of CDM and baryons relative to the one measured in the \GADGET{3} reference runs at $z=1$ for $\sum m_\nu = 0.15\,\rm{eV}$ for the simulations with a larger volume, $L_\mathrm{box} = 1024\,h^{-1}\,\rm{Mpc}$ and $N_\mathrm{part} = 1024^3$ (left panel), and at a higher resolution in the small volume, $L_\mathrm{box} = 512\,h^{-1}\,\rm{Mpc}$ and $N_\mathrm{part} = 1024^3$ (right panel). Results shown are for the squeezed configuration where $k_1 = k_2 \equiv k$, $k_3 = 0.012 \,h\,\mathrm{Mpc}^{-1}$. The grey bands highlight the interval of $\pm\,0.01$.}
  \label{fig:squeezedLBHR}
\end{figure}

As with the case of the power spectrum, we conducted various checks concerning numerical convergence with respect to finite-volume and resolution effects. These show a consistent picture that is in line with what we discussed in Section~\ref{subsec:convergencePS}. As an example, Figure~\ref{fig:squeezedLBHR} presents the results for the case of the squeezed configuration in simulations with $N_\mathrm{part} = 1024^3$, with a larger volume (left panel) or a higher resolution (right panel) than our simulations with $N_\mathrm{part} = 512^3$. In both cases, the agreement on smaller scales is improved: for the larger volume this happens because more independent triangles contribute to each measurement, while for the higher resolution this is due to the better numerical convergence of the density field on small scales.

Overall we may conclude that the $N$-body methods which produce highly consistent two-point statistics also tend to agree very well on the three-point statistics presented here. \NMGADGET appears to be an outlier, showing considerable deviations for squeezed configurations when the sum of the neutrino masses is large. This is however not unexpected since the final gauge transformation from Newtonian motion gauge has been neglected here. This transformation mainly acts at large scales and would therefore affect the squeezed configurations. At low neutrino mass, where the method works best, this effect is almost negligible though. It also becomes minimal at redshift $z = 0$ which was set as the target redshift for this method.

\subsection{Halo mass function}

\begin{figure}
  \includegraphics[width=\linewidth]{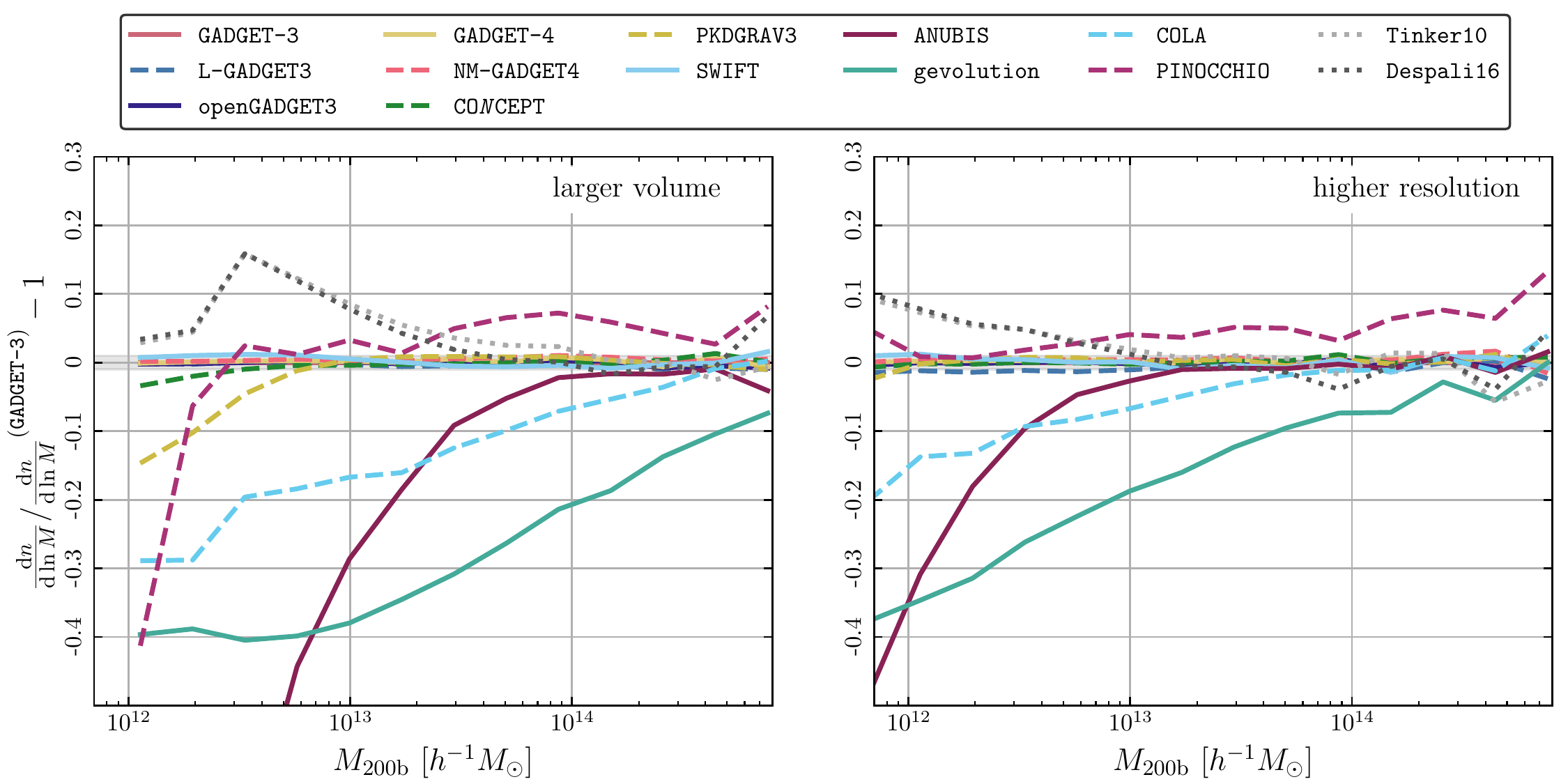}
  \caption{Halo mass functions relative to the one of \GADGET{3}, for $N_\mathrm{part} = 1024^3$ at $z = 0$ and neutrino mass $\sum m_\nu = 0.15$\,eV. The result for the larger-volume runs with $L_\mathrm{box} = 1024\,h^{-1}\,\rm{Mpc}$ is shown in the left panel while the right panel shows the result for the higher-resolution runs with $L_\mathrm{box} = 512\,h^{-1}\,\rm{Mpc}$.}
  \label{fig:hmf1}
\end{figure}

From the halo catalogues produced by \texttt{Denhf}, we estimate the halo mass functions (considering only the contribution from CDM and baryons) and compare them to the predictions by Tinker et al.~\cite{Tinker:2010my}, hereafter Tinker10, as well as Despali et al.~\cite{Despali:2015yla}, hereafter Despali16. For these predictions, we use the linear power spectra of CDM and baryons calculated by \CLASS for the respective neutrino cosmologies in the modelling of the theoretical halo mass functions. It has been shown by Costanzi et al.~\cite{Costanzi:2013bha} that this approach reproduces the halo mass function well for neutrino cosmologies. Figure~\ref{fig:hmf1} shows the ratio of the halo mass functions relative to the halo mass function of \GADGET{3} at a neutrino mass of $\sum m_\nu  = 0.15\,\rm{eV}$ for the runs with $N_\mathrm{part} = 1024^3$, in the large volume where $L_\mathrm{box} = 1024\,\si{\hMpc}$ (left panel), as well as for the higher-resolution setup with $L_\mathrm{box} = 512\,\si{\hMpc}$ (right panel). At the high-mass end, the agreement between different codes is generally very good, and fluctuations are smaller in the larger volume due to better statistics. At low masses $M_\mathrm{200b} < 10^{13}\,h^{-1}\si{\solarmass}$, the number density of halos is underestimated by \COLA, \gevolution, and \ANUBIS by up to $50 \%$ for the lower resolution. At higher resolution the agreement improves. The relatively poor performance of \gevolution in predicting the halo mass function can be understood from the fact that the code uses a uniform mesh. This leads to a smoothing of small-scale structures and generally to a mass estimate of halos that is poorly converged at the low-mass end. \COLA suffers from the same limitation, but the simulations used a mesh with significantly higher resolution in this case. Specifically, in the runs with $N_\mathrm{part} = 1024^3$, \COLA used a mesh of $3072^3$ grid points, while \gevolution used a mesh of $2048^3$ grid points.\footnote{Running \gevolution on such a fine mesh, i.e.\ with less than one particle per cell on average, was not supported in the public release of the code. In such a situation, the evolution would become unstable due to the low order of the finite-difference gradients used in the particle update. A second-order gradient computation was therefore implemented for this work, a feature that is made available in a recent patch of the code.}

\begin{figure}
  \includegraphics[width=\linewidth]{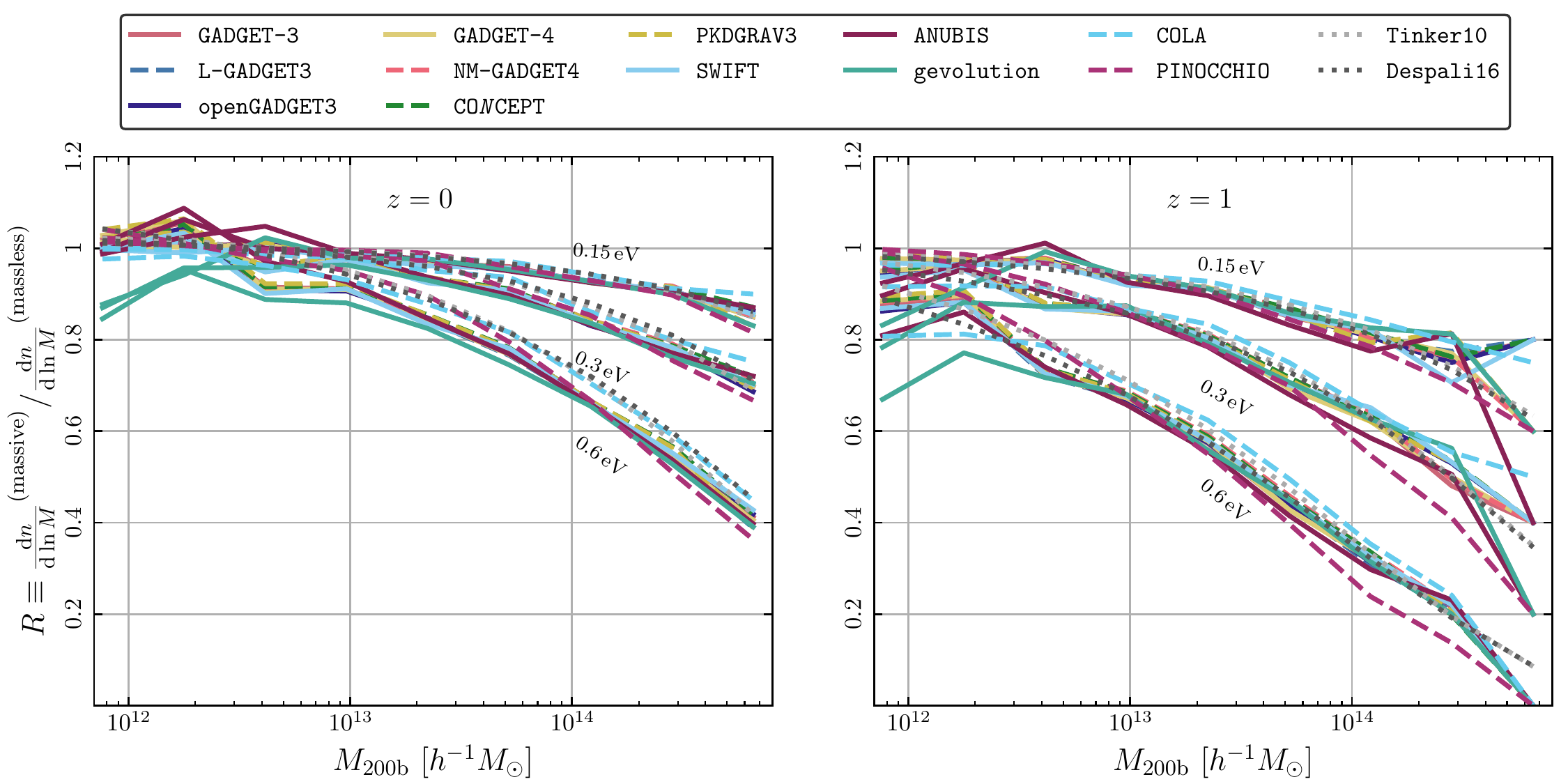}
  \caption{The halo mass function for different neutrino masses, relative to the case with massless neutrinos for each code, respectively. We show the results for the simulations with $L_\mathrm{box} = 512\,h^{-1}\,\rm{Mpc}$ and $N_\mathrm{part} = 512^3$ at $z = 0$ (left panel) and at $z = 1$ (right panel).}
  \label{fig:hmf2}
\end{figure}

\begin{figure}
  \includegraphics[width=\linewidth]{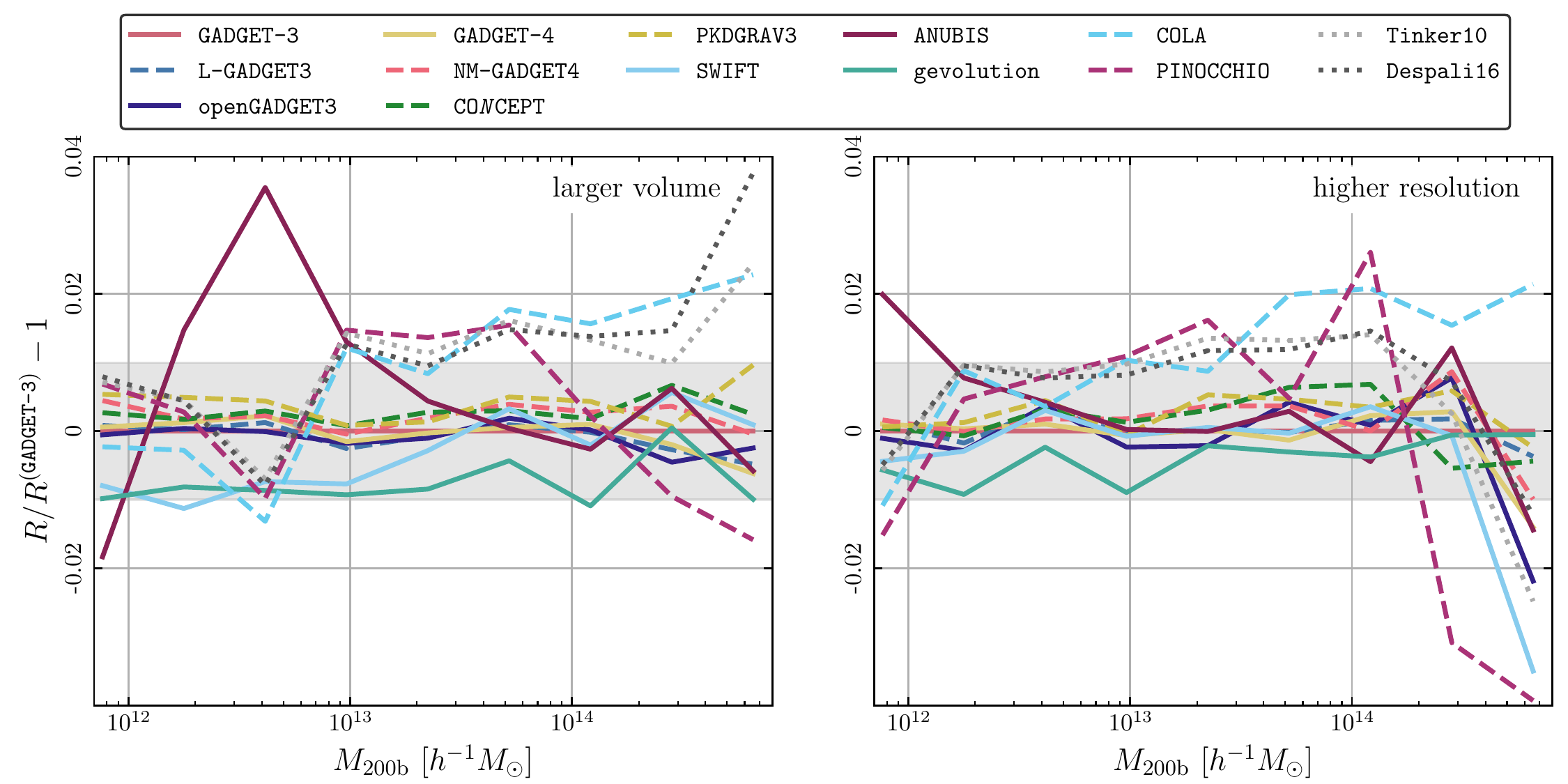}
   \caption{Suppression of the halo mass functions relative to the one measured in \GADGET{3} at redshift $z = 0$ and for a total neutrino mass of $\sum m_\nu = 0.15$\,eV. We show the results for the simulations with $L_\mathrm{box} = 1024\,h^{-1}\,\rm{Mpc}$ and $N_\mathrm{part} = 1024^3$ in the left panel, and $L_\mathrm{box} = 512\,h^{-1}\,\rm{Mpc}$ and $N_\mathrm{part} = 1024^3$ in the right panel. The grey bands highlight the interval of $\pm\,0.01$.}
  \label{fig:hmf3}
\end{figure}

Figure~\ref{fig:hmf2} shows the suppression of the halo mass function due to neutrinos with masses $\sum m_\nu \in \{0.15,\,0.3,\,0.6\}\,\rm{eV}$ at redshift $z = 0$ (left panel) and $z = 1$ (right panel). At low halo masses, there is little suppression, while going to higher masses the number density of halos is more and more suppressed. The higher the neutrino masses, and the higher the redshift, the stronger the suppression: at $z=0$ and $\sum m_\nu = 0.15\,\rm{eV}$ the suppression goes down to a factor of $0.9$ at halo masses of $10^{14}\,h^{-1}\si{\solarmass}$, while at $z=1$ and $\sum m_\nu = 0.6\,\rm{eV}$ the suppression goes down to $0.4$ at the same halo mass. In analogy to the case of the power spectrum, we define the suppression ratio with respect to the massless case as
\begin{equation}
    R = \frac{~\frac{\mathrm{d}n^\mathrm{(massive)}}{\mathrm{d}\ln M}~}{\frac{\mathrm{d}n^\mathrm{(massless)}}{\mathrm{d}\ln M}}\,.
\end{equation}
Figure~\ref{fig:hmf3} shows this suppression ratio relative to the one measured from \GADGET{3} for a sum of neutrino masses of $\sum m_\nu = 0.15$\,eV. The different codes generally agree within $3\%$ on the suppression ratio, even in cases where the halo mass function was poorly converged in Figure~\ref{fig:hmf1}. Our data show a trend that the suppression is about $1\%$ stronger than predicted by the models of Tinker10 and Despali16. The \COLA results are a slight outlier, agreeing more with these models than with the other simulations.

\subsection{Halo bias}

Finally, we study the halo bias for a fixed selection of halos defined by a mass threshold of $M_\mathrm{200b} > 10^{13}\,h^{-1}\si{\solarmass}$. However, since the halo mass function shows considerable differences between the different simulations, sometimes due to the fact that the mass estimate is not well converged at the low-mass end (certainly for \gevolution, \COLA, and \ANUBIS), we apply the halo selection as follows. First, we select the halos above the mass threshold in the reference runs done with \GADGET{3}. We may call the size of the selected population $N_\mathrm{h}$. Then, for each other code, we generate the sample by selecting the $N_\mathrm{h}$ most massive halos. The reasoning for this approach is that, while the estimated masses of individual halos may differ significantly between different codes, we still expect there to be a tight correlation that largely preserves the mass ordering. Another way to think about this is to consider a simple abundance matching of $N_\mathrm{h}$ sources, assigned to the centers of the most massive halos. We compare the bias measurements to the prediction by Tinker et al.~\cite{Tinker:2010my} (Tinker10). Here the large-scale bias is estimated from the mass-dependent peak height of halos in the linear density field. Given the \GADGET{3} halo masses, we model the peak heights using the linear power spectrum of CDM and baryons calculated by \CLASS for the respective neutrino cosmology. We then take the average of all biases obtained for each halo mass to get the final prediction. Note that the prediction of Tinker10 is only expected to work in the linear regime.

\begin{figure}
  \includegraphics[width=\textwidth]{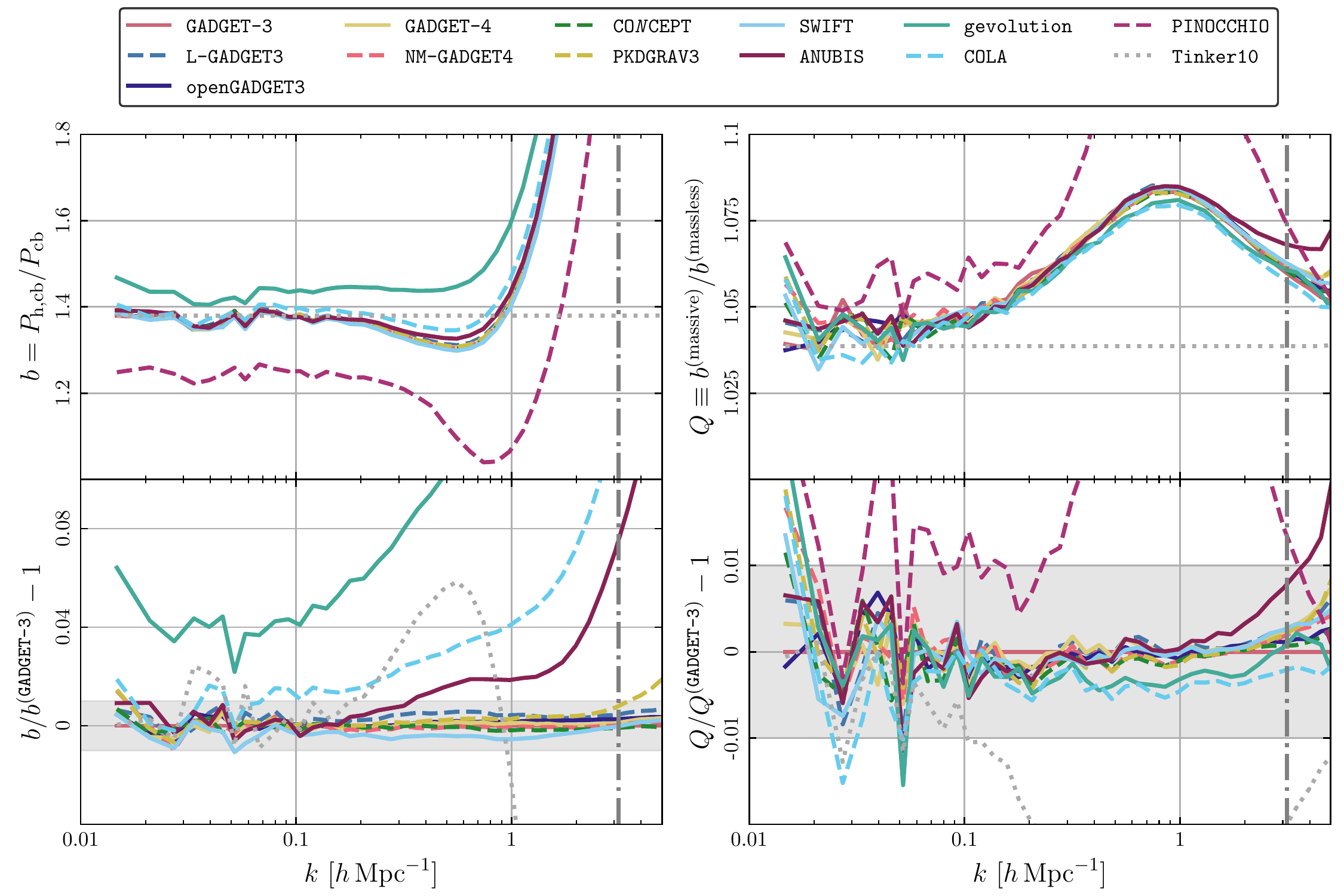}
  \caption{Halo bias with respect to CDM and baryons at $z = 0$ and neutrino mass $\sum m_\nu = 0.15$\,eV for the simulations with $L_\mathrm{box} = 1024\,h^{-1}\,\rm{Mpc}$ and $N_\mathrm{part} = 1024^3$. In the \GADGET{3} reference simulations, all halos with $M_\mathrm{200b} > 10^{13}\,h^{-1}\si{\solarmass}$ are selected, providing a sample of size $N_\mathrm{h}$. For the other simulations, we then select the most massive $N_\mathrm{h}$ halos. The grey bands in the lower panels highlight the interval of $\pm\,0.01$.}
  \label{fig:bias1}
\end{figure}

Figure~\ref{fig:bias1} shows the bias measurements from the simulations with larger volume, $L_\mathrm{box} = 1024\,h^{-1}\,\rm{Mpc}$. We define the scale-dependent halo bias $b(k)$ as the ratio of the cross-power spectrum of halos with CDM and baryons and the auto-power spectrum of CDM and baryons, i.e.
\begin{equation}
    b(k) = \frac{P_\mathrm{h,cb}(k)}{P_\mathrm{cb}(k)}\,.
\end{equation}
It has been shown by Castorina et al.~\cite{Castorina:2013wga} that defining the bias factor with respect to cold species gives closer-to-universal and less scale-dependent results than using the same definition with respect to total matter. As can be seen in the left panel of Figure~\ref{fig:bias1}, the bias measurements agree reasonably well on large scales except for \gevolution where the bias is measured to be about $4\%$ larger, and for \PINOCCHIO where the bias is measured to be about $10\%$ smaller. In analogy to the case of the power spectrum, we again define a bias ratio with respect to the massless case, which in this situation will quantify the increase (rather than suppression) of the bias in the presence of massive neutrinos,
\begin{equation}
    Q(k) = \frac{b^\mathrm{(massive)}(k)}{b^\mathrm{(massless)}(k)}\,.
\end{equation}
Results for this bias ratio are shown in the right panel of Figure~\ref{fig:bias1}. Here the agreement between different codes is excellent, well within $1\%$ over almost the entire range of scales probed. For \PINOCCHIO the bias ratio is about $1\%$ accurate up to $k \simeq 0.3\,h\,\mathrm{Mpc}^{-1}$.

\begin{figure}
  \includegraphics[width=\textwidth]{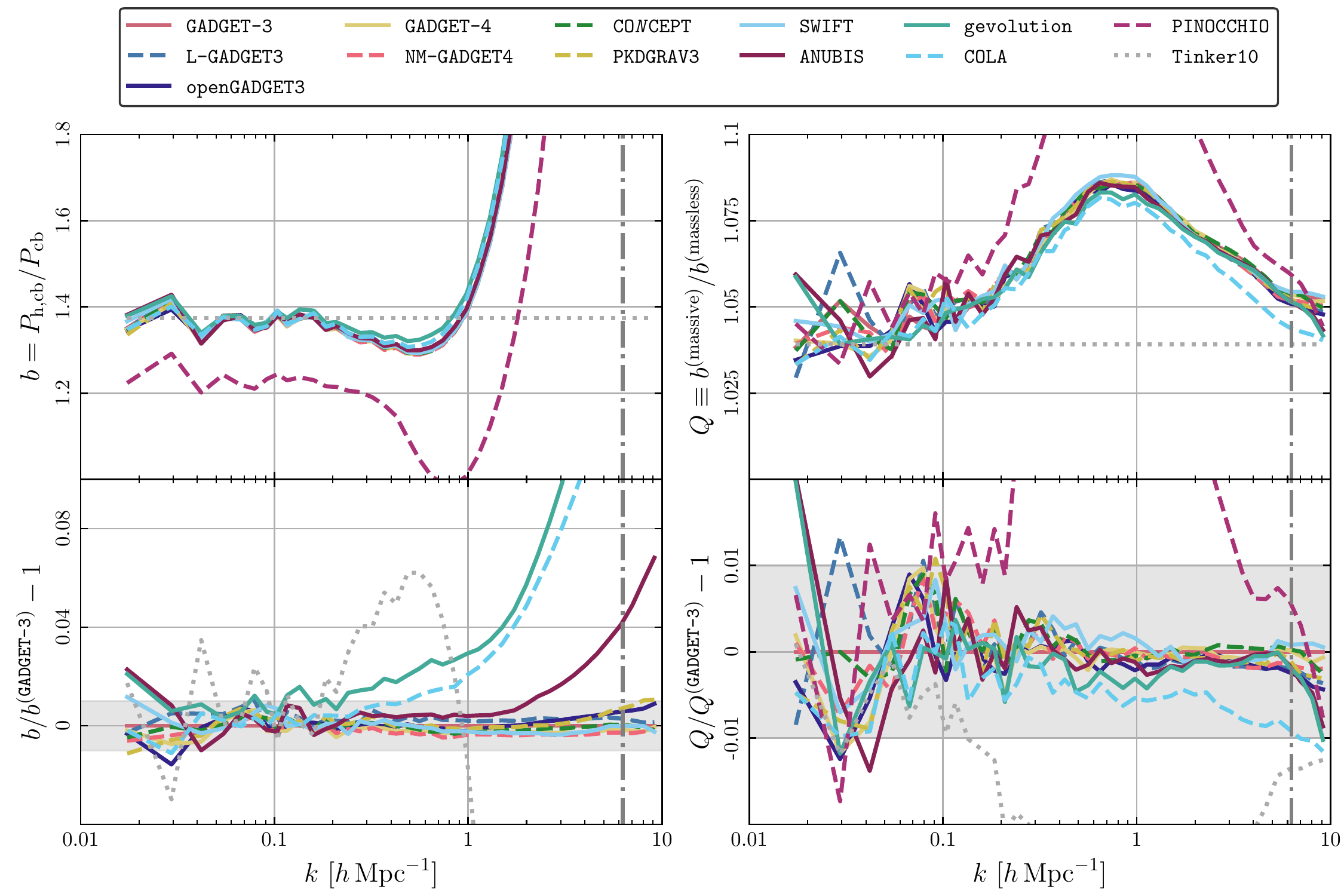}
  \caption{Halo bias with respect to CDM and baryons at $z = 0$ and neutrino mass $\sum m_\nu = 0.15$\,eV for the higher-resolution simulations with $L_\mathrm{box} = 512\,h^{-1}\,\rm{Mpc}$ and $N_\mathrm{part} = 1024^3$. In the \GADGET{3} reference simulations, all halos with $M_\mathrm{200b} > 10^{13}\,h^{-1}\si{\solarmass}$ are selected, providing a sample of size $N_\mathrm{h}$. For the other simulations, we then select the most massive $N_\mathrm{h}$ halos. The grey bands in the lower panels highlight the interval of $\pm\,0.01$.}
  \label{fig:bias2}
\end{figure}

To study the robustness of our results with respect to the mass resolution of the simulations we repeat the bias measurements in the runs with higher resolution, i.e.\ with $L_\mathrm{box} = 512\,\si{\hMpc}$ and $N_\mathrm{part} = 1024^3$. The smaller simulation volume leads to a higher level of shot noise in the halo counts, which incurs somewhat larger fluctuations when compared to the larger volume. As can be seen in Figure~\ref{fig:bias2}, the agreement between the different codes is improved significantly, in particular for codes that have difficulties in predicting the halo mass function accurately (\gevolution, \COLA, and \ANUBIS). The results for \PINOCCHIO do not improve significantly, as the discrepancy is mainly due to the approximate nature of the method rather than lack of resolution.

It is worth pointing out that the bias ratio $Q$ has a shape similar to the inverted power spectrum ratio $S_\mathrm{cb}$. This is actually expected, and was recently discussed by Hassani et al.~\cite{Hassani:2022yuq}. It means that the power spectrum of halos, when selected at fixed mass threshold, is much less sensitive to the neutrino mass than the power spectrum of CDM and baryons. On the other hand, synthetic catalogues are often created in such a way that the observed abundance of a certain type of object is reproduced (abundance matching). In such a situation, it may be more appropriate to study the dependence of the bias on the neutrino mass at fixed number count. We therefore repeat our measurements, but keeping the size of all samples fixed at $N_\mathrm{h}^\mathrm{(massless)}$, which is the number of halos with $M_\mathrm{200b} > 10^{13}\,h^{-1}\si{\solarmass}$ in the \GADGET{3} reference simulation at zero neutrino mass. In other words, when selecting the halo sample for non-zero neutrino mass, we still select the $N_\mathrm{h}^\mathrm{(massless)}$ most massive halos in all simulations. This effectively reduces the mass threshold of the selection for the massive neutrino case when compared to the previous procedure.

\begin{figure}
  \includegraphics[width=\textwidth]{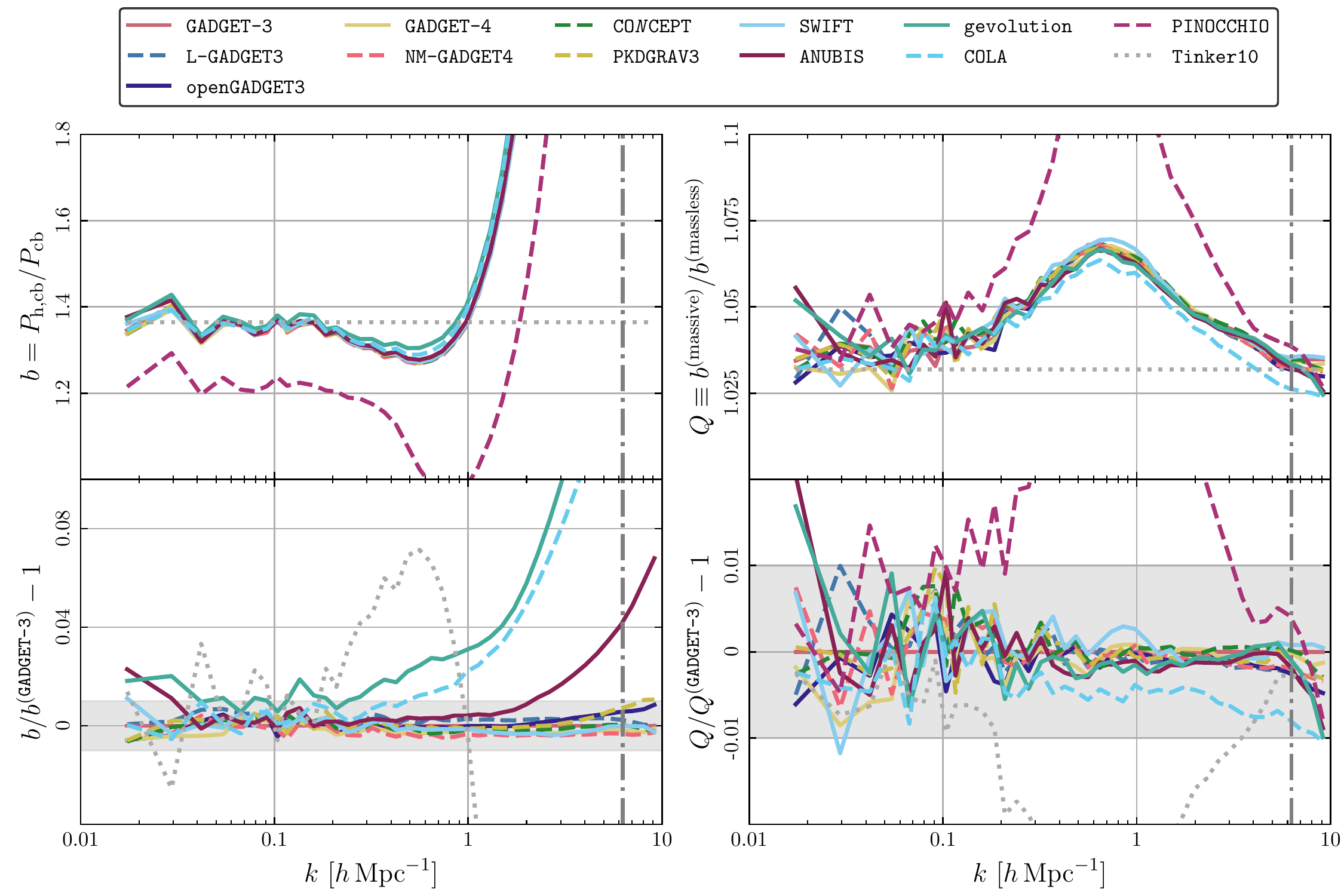}
  \caption{Halo bias with respect to CDM and baryons at $z = 0$ and neutrino mass $\sum m_\nu = 0.15$\,eV for the higher-resolution simulations with $L_\mathrm{box} = 512\,h^{-1}\,\rm{Mpc}$ and $N_\mathrm{part} = 1024^3$. In the \GADGET{3} reference simulation for $\sum m_\nu = 0$\,eV, all halos with $M_\mathrm{200b} > 10^{13}\,h^{-1}\si{\solarmass}$ are selected, providing a sample of size $N_\mathrm{h}^\mathrm{(massless)}$. We then select the most massive $N_\mathrm{h}^\mathrm{(massless)}$ halos in all the other simulations, even for those with non-zero neutrino mass. The grey bands in the lower panels highlight the interval of $\pm\,0.01$.}
  \label{fig:bias2_alt}
\end{figure}

Figure~\ref{fig:bias2_alt} shows the results of the bias measurement obtained through this procedure, using the higher-resolution simulations. We observe that the bias still increases with neutrino mass, but not quite as much as in Figure~\ref{fig:bias2} where a fixed halo mass threshold was used. The large-scale bias indicated by the dotted line (Tinker10) is about $0.01$, or one percentage point, lower so that the corresponding bias ratio $Q$ drops by $0.7\%$. The largest effect in $Q$ appears around $k \approx 1\,h\,\mathrm{Mpc}^{-1}$ were the change can be as much as $2\%$ due to the different halo selection. The interplay between bias enhancement and the suppression of the matter power spectrum has another layer of complexity due to the way in which the sample is selected. This needs to be studied carefully in the context of the specific numerical recipes that are employed in the production of synthetic catalogues.

\section{Discussion}
\label{sec:discussion}

Accurate and reliable modelling of the signatures that the neutrino mass imprints on the observables used to test the cosmological model is an essential ingredient for the data analysis of all upcoming galaxy surveys, and in particular for \textit{Euclid}. Such modelling necessarily requires a self-consistent description of the linear and possibly nonlinear clustering of neutrinos along with the nonlinear evolution of dark matter and baryonic structures. 
By comparing results across different implementations, including eleven full $N$-body implementations, two $N$-body schemes with fast time integration based on Lagrangian perturbation theory, and a further four codes that predict the nonlinear matter power spectra directly, we establish that current numerical techniques are in sub-percent agreement with regards to modelling the impact of massive neutrinos on the most common summary statistics of cosmological large-scale structure. We identify several specific situations where larger modelling errors can occur, but such shortfalls are generally well understood in terms of approximations or other compromises that were made in these situations. Our results can therefore be used as a detailed guide for choosing the preferred modelling techniques for any application given its requirements in terms of resources, accuracy, and quantities that need to be modelled. The validation presented here is the crucial first step for building up confidence in the numerical tools employed in the data analysis pipeline of \textit{Euclid}. It is particularly vital when considering the actual measurement of the neutrino mass scale from the data, which is one of the key science goals of the mission.

The fastest method of predicting simple summary statistics like the nonlinear matter power spectrum are emulators, and they will therefore play a crucial role in the cosmological likelihood analysis of \textit{Euclid} and other large-scale structure surveys. They are of course many orders of magnitude faster than simulations but tend to outperform even semi-analytic models which often have some bottlenecks in their numerical evaluation. However, emulators can only be as accurate as the simulations they are trained on, and it is therefore important to understand the modelling errors of simulations too. We find that the current state of the art for emulators yields an absolute precision on the power spectrum of total matter better than $2\%$ and can predict the relative change due to the neutrino mass to better than $1\%$ on all scales considered in this work. Interestingly, the best semi-analytic fitting methods available, in particular \ReACT and \HMcode, can achieve similar performance.

Overall our results demonstrate that we are in a fairly comfortable position, with several independent numerical techniques at our disposal that produce consistent results at the sub-percent level if we employ them diligently. The community has also implemented such techniques in a large number of different $N$-body codes, such that there is no shortage in choice of which code one wants to use. Moreover, our detailed comparison of particle-based and mesh-based techniques shows that the assumption of linear neutrinos is clearly sufficient to reach percent accuracy, even up to scales of the order of $k \approx 7\,h\,\mathrm{Mpc}^{-1}$ relevant for predicting the weak-lensing signal in \textit{Euclid}. We note that some codes have inherent difficulties reaching such levels of absolute accuracy due to effects of finite resolution. This is obvious in the cases where a uniform mesh is employed in the computation of gravitational interactions (\gevolution and \COLA), but AMR does not solve the issue entirely as the example of \ANUBIS illustrates. The relative impact of massive neutrinos can nonetheless be predicted very accurately with those codes. Here we do of course not attempt to address the additional challenge of modelling baryonic effects, i.e.\ astrophysical processes, down to such scales as this can be treated separately, see e.g.\ Martinelli et al.~\cite{Euclid:2020tff} for a discussion. On mildly nonlinear scales and in particular at redshifts $z \gtrsim 1$ relevant for \textit{Euclid}, a ``sophisticated 3LPT" realisation like the one produced with \PINOCCHIO, based on a scale-dependent linear growth rate computed from \CAMB and propagated to second- and third-order LPT with standard techniques, can be useful to produce a large number of halo catalogues in a limited amount of computing time, see e.g.\ Fumagalli et al.\ \cite{Euclid:2021api}.

The summary statistics considered in our analysis include auto- and cross-power spectra of the CDM and baryon component and neutrinos, bispectra of the CDM and baryon component, halo mass functions, and halo bias. We present results for redshifts $z=0$ and $z=1$, relevant for galaxy surveys like \textit{Euclid}. We do not consider redshift-space distortions or other effects that occur due to taking observations on our past light cone, and leave a detailed investigation of these effects to future work. However, we expect that no big surprises would appear given our level of confidence in the modelling of the summary statistics presented here.

In order to aid future code development, we make our reference simulations and analysis pipelines available via a public repository (see data availability statement below). This provides a reliable baseline against which further numerical methods can be validated, and it showcases the current state-of-the-art in modelling massive neutrinos in cosmology.

\acknowledgments
JA acknowledges financial support from the Swiss National Science Foundation. 
BB was supported by a UK Research and Innovation Stephen Hawking Fellowship (EP/W005654/1). 
MBa acknowledges support by the project ``Combining Cosmic Microwave Background and Large Scale Structure data: an Integrated Approach for Addressing Fundamental Questions in Cosmology", funded by the MIUR Progetti di Ricerca di Rilevante Interesse Nazionale (PRIN) Bando 2017 - grant 2017YJYZAH.
CH-A acknowledges support from the Excellence Cluster ORIGINS which is funded by the Deutsche Forschungsgemeinschaft (DFG, German Research Foundation) under Germany's Excellence Strategy - EXC-2094 - 390783311. 
BSW is supported by a Royal Society Enhancement Award (grant no.~RGF$\backslash$EA$\backslash$181023).
RM, DFM, HAW, FH thank the Research Council of Norway for their support and our computations were performed on resources provided by UNINETT Sigma2 -- the National Infrastructure for High Performance Computing and Data Storage in Norway.
CM acknowledges support from a UK Research and Innovation Future Leaders Fellowship [grant MR/S016066/1]. The \PINOCCHIO simulations were performed on Queen Mary's Apocrita HPC facility, supported by QMUL Research-IT.
CG acknowledges the support from the grant PRIN-MIUR 2017 WSCC32 ZOOMING and from the Italian National Institute of Astrophysics under the grant ``Bando PRIN 2019,'' 
PI: Viola Allevato.
TC is supported by the INFN INDARK PD51 grant and the FARE MIUR grant `ClustersXEuclid' R165SBKTMA. 
KD acknowledges support by the Deutsche Forschungsgemeinschaft (DFG, German Research Foundation) under Germany's Excellence Strategy - EXC-2094 - 390783311 and the COMPLEX project from the European Research Council (ERC) under the European Union's Horizon 2020 research and innovation program grant agreement ERC-2019-AdG 882679. The \openGADGET simulations were carried out at the Leibniz Supercomputer Center (LRZ) under the project pr86re. CA and BL are supported by an European Research Council Starting Grant (ERC-StG-716532), and BL also acknowledges support by the UK Science and Technology Facilities Council Consolidated Grants No.~ST/I00162X/1 and ST/P000541/1.
KK is supported by the UK Science and Technology Facilities Council (grant numbers ST/S000550/1 and ST/W001225/1). MV and GP are supported by the INFN INDARK PD 51 grant and by the 
ASI-INAF n.\ 2017-14-H.0 agreement. REA and MZ acknowledge the support of the ERC-STG number 716151 (BACCO).
This work used the DiRAC@Durham facility managed by the Institute for Computational Cosmology on behalf of the STFC DiRAC HPC Facility (\url{www.dirac.ac.uk}). The equipment was funded by BEIS via STFC capital grants ST/K00042X/1, ST/P002293/1, ST/R002371/1 and ST/S002502/1, Durham University and STFC operation grant ST/R000832/1. DiRAC is part of the UK National e-Infrastructure.

\AckEC

For the purpose of open access, the authors have applied a Creative Commons Attribution (CC BY) licence to any Author Accepted Manuscript version arising from this work and upload the accepted version to \texttt{arXiv} with CC BY licence. This is the Accepted Manuscript version of an article accepted for publication in JCAP. IOP Publishing Ltd is not responsible for any errors or omissions in this version of the manuscript or any version derived from it. The Version of Record is available online at \url{https://doi.org/10.1088/1475-7516/2023/06/035}.

\section*{Data availability}

Raw data from our reference runs (\GADGET{3}), and all reduced data (summary statistics) presented in this work are available at \url{https://doi.org/10.5281/zenodo.7868793}. We also provide initial data, parameter specifications for \CLASS, and a documented analysis pipeline to facilitate further validations against our results.

\bibliographystyle{utcaps}

\typeout{}
\bibliography{references}

\end{document}